\documentclass[12pt]{article}
\usepackage{amsmath,amssymb,amsthm,amsxtra,overpic,bbm,bm,epsfig,subfigure}
\usepackage{mathrsfs}
\usepackage{hyperref}
\usepackage{enumitem}
\usepackage{graphicx}
\usepackage{color}
\usepackage{comment}
\usepackage{epstopdf}
\usepackage{float}
\usepackage{cite}
\usepackage{slashed,stmaryrd}
\numberwithin{equation}{section}

\def\thefootnote{\fnsymbol{footnote}}

\begin{document}

\vspace{0.2cm}

\begin{center}
{\Large\bf Continuous and Discrete Symmetries of Renormalization Group Equations for Neutrino Oscillations in Matter}
\end{center}

\vspace{0.2cm}

\begin{center}
{\bf Shun Zhou}~\footnote{E-mail: zhoush@ihep.ac.cn}
\\
\vspace{0.2cm}
{\small
Institute of High Energy Physics, Chinese Academy of Sciences, Beijing 100049, China\\
School of Physical Sciences, University of Chinese Academy of Sciences, Beijing 100049, China}
\end{center}

\vspace{1.5cm}

\begin{abstract}
Three-flavor neutrino oscillations in matter can be described by three effective neutrino masses $\widetilde{m}^{}_i$ (for $i = 1, 2, 3$) and the effective mixing matrix $V^{}_{\alpha i}$ (for $\alpha = e, \mu, \tau$ and $i = 1, 2, 3$). When the matter parameter $a \equiv 2\sqrt{2} G^{}_{\rm F} N^{}_e E$ is taken as an independent variable, a complete set of first-order ordinary differential equations for $\widetilde{m}^2_i$ and $|V^{}_{\alpha i}|^2$ have been derived in the previous works. In the present paper, we point out that such a system of differential equations possesses both the continuous symmetries characterized by one-parameter Lie groups and the discrete symmetry associated with the permutations of three neutrino mass eigenstates. The implications of these symmetries for solving the differential equations and looking for differential invariants are discussed.
\end{abstract}


\def\thefootnote{\arabic{footnote}}
\setcounter{footnote}{0}
\newpage
\newpage

\section{Introduction}

In the past few decades, the phenomena of neutrino flavor conversions have been well established by a number of elegant neutrino oscillation experiments, providing us with a solid evidence that neutrinos are massive and lepton flavors are mixed~\cite{Xing:2011zza, Tanabashi:2018oca, Xing:2019vks}. When neutrinos are propagating a medium, the coherent forward scattering of neutrinos with background particles can significantly change the behaviors of neutrino flavor oscillations~\cite{Wolfenstein:1977ue, Mikheev:1986gs}. In the three-flavor framework, neutrino oscillations in ordinary matter are governed by the effective Hamiltonian~\cite{Barger:1980tf, Kuo:1989qe}
\begin{eqnarray}
H^{}_{\rm m} = \frac{1}{2E} \left[U \left(\begin{matrix} m^2_1 & 0 & 0 \cr 0 & m^2_2 & 0 \cr 0 & 0 & m^2_3  \end{matrix}\right) U^\dagger + \left( \begin{matrix} a & 0 & 0 \cr 0 & 0 & 0 \cr 0 & 0 & 0 \end{matrix} \right)\right] \equiv \frac{1}{2E} \left[ V \left( \begin{matrix} \widetilde{m}^2_1 & 0 & 0 \cr 0 & \widetilde{m}^2_2 & 0 \cr 0 & 0 & \widetilde{m}^2_3 \end{matrix}\right) V^\dagger\right]\; ,
\label{eq:Hm}
\end{eqnarray}
where $E$ stands for the neutrino energy, $U$ is the $3\times 3$ unitary matrix of lepton flavor mixing, and $m^{}_i$ (for $i = 1, 2, 3$) are neutrino masses in vacuum. In addition, the parameter $a \equiv 2\sqrt{2}G^{}_{\rm F} N^{}_e E$, with $G^{}_{\rm F}$ being the Fermi constant and $N^{}_e$ being the net number density of electrons, signifies the matter effects on neutrino oscillations. In the last step in Eq.~(\ref{eq:Hm}), the effective Hamiltonian $H^{}_{\rm m}$ has been diagonalized by the effective mixing matrix $V$ in matter, which is also a $3\times 3$ unitary matrix, and $\widetilde{m}^{}_i$ (for $i = 1, 2, 3$) in the eigenvalues denote the effective neutrino masses in matter.

Although it is always possible to explicitly figure out three eigenvalues and the corresponding eigenvectors of the effective Hamiltonian $H^{}_{\rm m}$ by solving the characteristic equation~\cite{Barger:1980tf, Zaglauer:1988gz, Krastev:1988yu, Xing:2000gg, Xing:2016ymg}, two interesting identities have been derived in the literature to set up direct connections between the fundamental parameters in vacuum and the effective ones in matter~\cite{Naumov:1991ju, Harrison:1999df, Kimura:2002hb, Kimura:2002wd}
\begin{eqnarray}
\frac{\widetilde{\cal J}}{\cal J} = \frac{\Delta^{}_{12} \Delta^{}_{23} \Delta^{}_{31}}{\widetilde{\Delta}^{}_{12} \widetilde{\Delta}^{}_{23} \widetilde{\Delta}^{}_{31}} =  \frac{|V^{}_{e1}|\cdot|V^{}_{e2}| \cdot |V^{}_{e3}| }{|U^{}_{e1}| \cdot |U^{}_{e2}| \cdot |U^{}_{e3}|} \; ,
\label{eq:identities}
\end{eqnarray}
where the neutrino mass-squared differences $\Delta^{}_{ij} \equiv m^2_i - m^2_j$ (for $ij = 12, 23, 31$) and the Jarlskog invariant ${\cal J} \equiv {\rm Im}(U^{}_{e1} U^*_{e2} U^*_{\mu 1} U^{}_{\mu 2})$ in vacuum have been defined~\cite{Jarlskog:1985ht, Wu:1985ea, Cheng:1986in}, and likewise for their counterparts $\widetilde{\Delta}^{}_{ij} \equiv \widetilde{m}^2_i - \widetilde{m}^2_j$ and $\widetilde{\cal J} \equiv {\rm Im}(V^{}_{e1} V^*_{e2} V^*_{\mu 1} V^{}_{\mu 2})$ in matter. The first identity in Eq.~(\ref{eq:identities}) is also referred to as the Naumov relation~\cite{Naumov:1991ju}. The second one together with the Naumov relation leads to the Toshev relation $\sin 2\widetilde{\theta}^{}_{23} \sin\widetilde{\delta} = \sin 2\theta^{}_{23} \sin\delta$~\cite{Toshev:1991ku}, if the standard parametrization~\cite{Tanabashi:2018oca} in terms of three mixing angles $\{\widetilde{\theta}^{}_{12}, \widetilde{\theta}^{}_{13}, \widetilde{\theta}^{}_{23}\}$ and one Dirac CP-violating phase $\widetilde{\delta}$ is adopted for the mixing matrix $V$ in matter, and similarly $\{\theta^{}_{12}, \theta^{}_{13}, \theta^{}_{23}\}$ and $\delta$ for the mixing matrix $U$ in vacuum.

Recently, a complete set of first-order ordinary differential equations for $\widetilde{m}^2_i$ (for $i = 1, 2, 3$) and $|V^{}_{\alpha i}|^2$ (for $\alpha = e, \mu, \tau$ and $i = 1, 2, 3$) have been obtained by differentiating the effective Hamiltonian in Eq.~(\ref{eq:Hm}) with respect to the matter parameter $a$~\cite{Chiu:2017ckv, Xing:2018lob}. A close analogy has been made in Ref.~\cite{Xing:2018lob} between these differential equations for the effective neutrino mixing parameters in matter and the renormalization-group equations (RGEs) for the running flavor mixing parameters~\cite{Xing:2018kto}. The approximate analytical solutions to these RGEs for neutrino oscillations in matter have been found~\cite{Wang:2019yfp} and applied to the studies of leptonic CP violation~\cite{Wang:2019dal, Petcov:2018zka}. In the present work, we attempt to further explore the practical advantages of the RGE approach to neutrino oscillations in matter and investigate whether one can gain more insights into matter effects on neutrino oscillations in this particularly useful language. Such an exploration is mainly motivated by the following questions.
\begin{itemize}
\item Do there exist any other identities independent of those two in Eq.~(\ref{eq:identities})? If the answer is affirmative, then how many identities of this kind are there? The answers to these questions are rather nontrivial. First, let us recapitulate those two independent identities in Eq.~(\ref{eq:identities}) in a more transparent way
    \begin{eqnarray}
    \widetilde{\cal J} \widetilde{\Delta}^{}_{12} \widetilde{\Delta}^{}_{23} \widetilde{\Delta}^{}_{31} &=& {\cal J} \Delta^{}_{12} \Delta^{}_{23} \Delta^{}_{31} \; , \label{eq:Naumov}\\
    |V^{}_{e1}| |V^{}_{e2}| |V^{}_{e3}| \widetilde{\Delta}^{}_{12} \widetilde{\Delta}^{}_{23} \widetilde{\Delta}^{}_{31} &=&
    |U^{}_{e1}| |U^{}_{e2}| |U^{}_{e3}| \Delta^{}_{12} \Delta^{}_{23} \Delta^{}_{31} \; ,
    \label{eq:Vei}
    \end{eqnarray}
where the effective and fundamental neutrino parameters are located on the left- and right-hand side, respectively. Note that there is no explicit dependence on the matter parameter $a$ on both sides. Second, only the parameters relevant for neutrino oscillations are present. In other words, the neutrino mass-squared differences $\widetilde{\Delta}^{}_{ij}$ and $\Delta^{}_{ij}$ instead of the squared neutrino masses $\widetilde{m}^2_i$ and $m^2_i$ themselves are involved.

\item In Ref.~\cite{Xing:2018lob}, it has been discovered that the RGEs for $\{|V^{}_{e1}|^2, |V^{}_{e2}|^2, |V^{}_{e3}|^2\}$ and $\{\widetilde{\Delta}^{}_{12}, \widetilde{\Delta}^{}_{23}, \widetilde{\Delta}^{}_{31}\}$ appear to be in a closed form, namely,
    \begin{eqnarray}
\frac{\rm d}{{\rm d}a} |V^{}_{e1}|^2 &=& 2 |V^{}_{e1}|^2 \left(|V^{}_{e2}|^2 \widetilde{\Delta}^{-1}_{12} - |V^{}_{e3}|^2 \widetilde{\Delta}^{-1}_{31}\right) \; ,
\label{eq:RGEe1}\\
\frac{\rm d}{{\rm d}a} |V^{}_{e2}|^2 &=& 2 |V^{}_{e2}|^2 \left(|V^{}_{e3}|^2 \widetilde{\Delta}^{-1}_{23} - |V^{}_{e1}|^2 \widetilde{\Delta}^{-1}_{12}\right) \; ,
\label{eq:RGEe2}\\
\frac{\rm d}{{\rm d}a} |V^{}_{e3}|^2 &=& 2 |V^{}_{e3}|^2 \left(|V^{}_{e1}|^2 \widetilde{\Delta}^{-1}_{31} - |V^{}_{e2}|^2 \widetilde{\Delta}^{-1}_{23}\right) \; ,
\label{eq:RGEe3}
\end{eqnarray}
and
\begin{eqnarray}
\frac{\rm d}{{\rm d}a} \widetilde{\Delta}^{}_{12} &=& |V^{}_{e1}|^2 - |V^{}_{e2}|^2 \; ,
\label{eq:Del12}\\
\frac{\rm d}{{\rm d}a} \widetilde{\Delta}^{}_{23} &=& |V^{}_{e2}|^2 - |V^{}_{e3}|^2 \; ,
\label{eq:Del23}\\
\frac{\rm d}{{\rm d}a} \widetilde{\Delta}^{}_{31} &=& |V^{}_{e3}|^2 - |V^{}_{e1}|^2 \; .
\label{eq:Del31}
\end{eqnarray}
Obviously, due to the normalization condition $|V^{}_{e1}|^2 + |V^{}_{e2}|^2 + |V^{}_{e3}|^2 = 1$ and the trivial identity $\widetilde{\Delta}^{}_{12} + \widetilde{\Delta}^{}_{23} + \widetilde{\Delta}^{}_{31} = 0$, we are left with only four independent equations. However, it has been pointed out in Ref.~\cite{Xing:2018lob} that the above array of six differential equations respect the permutation symmetry $S^{}_3$. Such a discrete symmetry simply originates from the effective Hamiltonian, which is invariant under the permutations of the labels of three neutrino mass eigenvalues $\{\widetilde{m}^{}_1, \widetilde{m}^{}_2, \widetilde{m}^{}_3\}$ and the corresponding eigenvectors $\{V^{}_{\alpha 1}, V^{}_{\alpha 2}, V^{}_{\alpha 3}\}$. See, Refs.~\cite{Kuo:2018mnm} and \cite{Kuo:2019psm}, for a recent discussion on the discrete symmetries of this type in flavor physics.

An immediate question is whether this discrete symmetry could help us to solve the RGEs or shed some light on the construction of useful identities similar to those in Eqs.~(\ref{eq:Naumov}) and (\ref{eq:Vei}). In Refs.~\cite{Chiu:2017ckv} and \cite{Xing:2018lob}, the identities in Eqs.~(\ref{eq:Naumov}) and (\ref{eq:Vei}) have been shown to be equivalent to differential invariants ${\cal I}^{}_1 = \ln(\widetilde{\cal J} \widetilde{\Delta}^{}_{12} \widetilde{\Delta}^{}_{23} \widetilde{\Delta}^{}_{31})$ and ${\cal I}^{}_2 = \ln(|V^{}_{e1}|^2 |V^{}_{e2}|^2 |V^{}_{e3}|^2 \widetilde{\Delta}^2_{12} \widetilde{\Delta}^2_{23} \widetilde{\Delta}^2_{31})$, for which ${\rm d}{\cal I}^{}_1/{\rm d}a = 0$ and ${\rm d}{\cal I}^{}_2/{\rm d}a = 0$ can be proved.
\end{itemize}
Bearing these questions in mind, we shall concentrate on the differential equations in Eqs.~(\ref{eq:RGEe1})-(\ref{eq:Del31}) and examine their possible symmetries, both continuous and discrete. As will be shown later, it is the continuous symmetries characterized by one-parameter Lie groups that play a very important role in solving the RGEs and understanding the relevant identities in Eqs.~(\ref{eq:Naumov}) and (\ref{eq:Vei}) as differential invariants.

The remaining part of this paper is organized as follows. In Sec.~\ref{sec:two}, we take the simple case of two-flavor neutrino oscillations in matter as an illustrative example. Some general theorems for the solutions to ordinary differential equations are introduced and applied to the two-flavor case. It turns out that two differential equations in this case are completely solvable by quadratures. Then, the three-flavor neutrino oscillations are discussed in Sec.~\ref{sec:three}, where we explain why the differential equations cannot be simply solved by quadratures alone and demonstrate how to reduce the number of differential equations with the help of the continuous symmetries characterized by two one-parameter Lie groups and the permutation symmetry $S^{}_3$. Some brief remarks on the direction of further developments are given in Sec.~\ref{sec:remark}. Then we summarize our main conclusions in Sec.~\ref{sec:sum}. Basic properties of the $S^{}_3$ group are collected in Appendix~\ref{sec:app}.

\section{Two-flavor Neutrino Oscillations}
\label{sec:two}

As a warm-up exercise, let us first consider two-flavor neutrino oscillations in matter, for which the effective Hamiltonian is given by
\begin{eqnarray}
H^{2\nu}_{\rm m} = \frac{1}{2E} \left[ U \left(\begin{matrix} m^2_1 & 0 \cr 0 & m^2_2 \end{matrix}\right) U^\dagger + \left(\begin{matrix} a & 0 \cr 0 & 0 \end{matrix}\right) \right] \equiv \frac{1}{2E} \left[ V \left(\begin{matrix} \widetilde{m}^2_1 & 0 \cr 0 & \widetilde{m}^2_2 \end{matrix}\right) V^\dagger \right]\; ,
\label{eq:Heff2nu}
\end{eqnarray}
where the notations follow exactly those in Eq.~(\ref{eq:Hm}) but now for only two lepton flavors. For definiteness, we specify the subscripts of the mixing matrix elements $U^{}_{\alpha i}$ and $V^{}_{\alpha i}$ as $\alpha = e, \mu$ and $i = 1, 2$. Differentiating Eq.~(\ref{eq:Heff2nu}) with respect to $a$, one can obtain~\cite{Chiu:2017ckv, Xing:2018lob}
\begin{eqnarray}
\frac{\rm d}{{\rm d}a} |V^{}_{e1}|^2 &=& 2 |V^{}_{e1}|^2 |V^{}_{e2}|^2 \widetilde{\Delta}^{-1}_{12} \; ,
\label{eq:RGE2nue1} \\
\frac{\rm d}{{\rm d}a} |V^{}_{e2}|^2 &=& 2 |V^{}_{e1}|^2 |V^{}_{e2}|^2 \widetilde{\Delta}^{-1}_{21} \; ,
\label{eq:RGE2nue2} \\
\frac{\rm d}{{\rm d}a} \widetilde{\Delta}^{}_{12} &=&  |V^{}_{e1}|^2 - |V^{}_{e2}|^2  \; ,
\label{eq:RGE2nuDel12}\\
\frac{\rm d}{{\rm d}a} \widetilde{\Delta}^{}_{21} &=&  |V^{}_{e2}|^2 - |V^{}_{e1}|^2  \; .
\label{eq:RGE2nuDel21}
\end{eqnarray}
It is evident that only two of the above first-order differential equations are independent due to the unitarity condition $|V^{}_{e1}|^2 + |V^{}_{e2}|^2 = 1$ and the trivial identity $\widetilde{\Delta}^{}_{12} + \widetilde{\Delta}^{}_{21} = 0$. However, we can immediately observe that Eqs.~(\ref{eq:RGE2nue2}) and (\ref{eq:RGE2nuDel21}) are respectively related to Eqs.~(\ref{eq:RGE2nue1}) and (\ref{eq:RGE2nuDel12}) via the $Z^{}_2$ symmetry generated by the $1 \leftrightarrow 2$ transposition.

If one simply chooses two independent quantities, such as $|V^{}_{e1}|^2$ and $\widetilde{\Delta}^{}_{21}$, then two independent differential equations in Eqs.~(\ref{eq:RGE2nue1}) and (\ref{eq:RGE2nuDel12}) can be rewritten as
\begin{eqnarray}
\frac{\rm d}{{\rm d}a} |V^{}_{e1}|^2 &=& - 2 |V^{}_{e1}|^2 ( 1 - |V^{}_{e1}|^2) \widetilde{\Delta}^{-1}_{21} \; ,
\label{eq:RGE2nue1n}\\
\frac{\rm d}{{\rm d}a} \widetilde{\Delta}^{}_{21} &=& 1 - 2 |V^{}_{e1}|^2 \; ,
\label{eq:RGE2nuDel12n}
\end{eqnarray}
where the $Z^{}_2$ symmetry is explicitly broken. As we have briefly mentioned, such a discrete symmetry originates from the invariance of the effective Hamiltonian in Eq.~(\ref{eq:Heff2nu}) under the permutation of neutrino mass eigenstates~\cite{Zhou:2016luk}. In order to preserve this $Z^{}_2$ symmetry, we introduce another two independent functions $|V^{}_{e1}|^2/\widetilde{\Delta}^{}_{21}$ and $|V^{}_{e2}|^2/\widetilde{\Delta}^{}_{21}$, and recast two differential equations into
\begin{eqnarray}
\frac{\rm d}{{\rm d}a} \left( \frac{|V^{}_{e1}|^2}{\widetilde{\Delta}^{}_{21}} \right) &=& - 2\frac{|V^{}_{e1}|^2}{\widetilde{\Delta}^{}_{21}} \cdot \frac{|V^{}_{e2}|^2}{\widetilde{\Delta}^{}_{21}} -  \frac{|V^{}_{e1}|^2}{\widetilde{\Delta}^{}_{21}} \left( \frac{|V^{}_{e2}|^2}{\widetilde{\Delta}^{}_{21}} - \frac{|V^{}_{e1}|^2}{\widetilde{\Delta}^{}_{21}} \right) \; ,
\label{eq:RGE2nusym1} \\
\frac{\rm d}{{\rm d}a} \left( \frac{|V^{}_{e2}|^2}{\widetilde{\Delta}^{}_{21}} \right) &=& +  2\frac{|V^{}_{e1}|^2}{\widetilde{\Delta}^{}_{21}} \cdot \frac{|V^{}_{e2}|^2}{\widetilde{\Delta}^{}_{21}} -  \frac{|V^{}_{e2}|^2}{\widetilde{\Delta}^{}_{21}} \left( \frac{|V^{}_{e2}|^2}{\widetilde{\Delta}^{}_{21}} - \frac{|V^{}_{e1}|^2}{\widetilde{\Delta}^{}_{21}} \right) \; .
\label{eq:RGE2nusym2}
\end{eqnarray}
Needless to say, these two equations are equivalent to those in Eqs.~(\ref{eq:RGE2nue1n}) and (\ref{eq:RGE2nuDel12n}), but in a more symmetric form. Moreover, these differential equations will be further simplified by defining two new functions
\begin{eqnarray}
x^{}_+ \equiv \frac{|V^{}_{e2}|^2 - |V^{}_{e1}|^2}{\widetilde{\Delta}^{}_{21}} \; , \quad x^{}_- \equiv \frac{|V^{}_{e2}|^2 + |V^{}_{e1}|^2}{\widetilde{\Delta}^{}_{21}}
\label{eq:xpm}
\end{eqnarray}
such that Eqs.~(\ref{eq:RGE2nusym1}) and (\ref{eq:RGE2nusym2}) can be converted into a very simple form
\begin{eqnarray}
\frac{{\rm d}x^{}_+}{{\rm d}a} &=& x^2_- - 2 x^2_+ \; , \label{eq:xpd} \\
\frac{{\rm d}x^{}_-}{{\rm d}a} &=& - x^{}_+ x^{}_- \; . \label{eq:xmd}
\end{eqnarray}
It is worthwhile to emphasize that $x^{}_+ \rightarrow + x^{}_+$ and $x^{}_- \rightarrow - x^{}_-$ hold under the exchange of neutrino mass indices $1 \leftrightarrow 2$, so they serve as the trivial and nontrivial one-dimensional representations of the $Z^{}_2$ symmetry, respectively. One can easily observe that Eqs.~(\ref{eq:xpd}) and (\ref{eq:xmd}) are indeed invariant under these transformations.

The primary task in this section is to explicitly solve Eqs.~(\ref{eq:xpd}) and (\ref{eq:xmd}) and find out the differential invariants. To this end, we implement the continuous symmetries of one-parameter Lie groups to reduce the number of first-order differential equations, following the strategy presented in the famous book by Peter J. Olver~\cite{Olver}. For completeness, we first sketch the main relevant ideas from this book and then apply them to the concrete problem.
\begin{itemize}
\item The manifold of our interest is actually the $m$-dimensional Euclidean space $\mathbb{R}^m$, and any point $x \in \mathbb{R}^m$ can be represented by $m$ local coordinates $x = (x^1, x^2, \dots, x^m)$. The vector field ${\bf v}|^{}_x$ defined on the manifold is written as ${\bf v}|^{}_x = \xi^1(x) \partial/\partial x^1 + \xi^2(x) \partial/\partial x^2 + \cdots + \xi^m(x) \partial/\partial x^m$, where $\xi^i(x)$ for $i = 1, 2, \dots, m$ are all smooth functions. The action of a local Lie group $G$ on the manifold can be generated by the exponential maps of all the vector fields ${\bf v}$'s that constitute the corresponding Lie algebra. The system of $q$ differential equations up to the order $n$ is denoted as $\Delta^{}_\nu(x,u^{(n)}) = 0$ for $\nu = 1, 2, \dots, q$, where $x = (x^1, x^2, \dots, x^r)$ stand for $r$ independent variables and $u^{(n)}$ for $s$ dependent variables. Then, we focus on this $(r+s)$-dimensional smooth manifold and investigate the Lie groups of transformations that leave the differential equations invariant, implying that the solutions to the differential equations have been transformed from one to another. According to {\bf Theorem 2.68} in Ref.~\cite{Olver}, if $q$ first-order differential equations are invariant under a $q$-dimensional Lie group, then these differential equations can be solved by quadratures alone. Therefore, we have to see whether Eqs.~(\ref{eq:xpd}) and (\ref{eq:xmd}) possess some continuous symmetries associated with a two-dimensional Lie group.

\item It is obvious that this system is invariant under the group $G^{}_1$ of translations $(a, x^{}_+, x^{}_-) \mapsto (a + \varepsilon, x^{}_+, x^{}_-)$ in $\mathbb{R}^3$, where $\varepsilon$ is a real constant. This one-parameter Lie group is generated by the vector field ${\bf v}^{}_1 = \partial/\partial a \equiv \partial^{}_a$. In addition, the scaling transformations $(a, x^{}_+, x^{}_-) \mapsto (\lambda^{-1} a, \lambda x^{}_+, \lambda x^{}_-)$ with $\lambda \neq 0$ leave Eqs.~(\ref{eq:xpd}) and (\ref{eq:xmd}) unchanged. It is straightforward to figure out the generator of this one-parameter Lie group $G^{}_2$ as ${\bf v}^{}_2 = -a\partial^{}_a + x^{}_+ \partial^{}_+ + x^{}_-\partial^{}_-$, where $\partial^{}_+ \equiv \partial/\partial x^{}_+$ and $\partial^{}_- \equiv \partial/\partial x^{}_-$ have been defined. Since these two generators span a two-dimensional Lie algebra, namely,
\begin{eqnarray}
[{\bf v}^{}_1, {\bf v}^{}_2] = -{\bf v}^{}_1 \; ,
\label{eq:algebra}
\end{eqnarray}
the corresponding Lie group is solvable. As a consequence, Eqs.~(\ref{eq:xpd}) and (\ref{eq:xmd}) can be completely solved by quadratures.

\item Then we proceed to explicitly solve these two differential equations. For the translation group generated by ${\bf v}^{}_1 = \partial^{}_a$, one usually needs to find out a new system of coordinates $y = y(a, x^{}_+, x^{}_-)$, $u = u(a, x^{}_+, x^{}_-)$ and $w = w(a, x^{}_+, x^{}_-)$ such that ${\bf v}^{}_1(y) = 0$, ${\bf v}^{}_1 (u) = 0$ and ${\bf v}^{}_1(w) = 1$. The vector field itself in these new coordinates becomes ${\bf v}^{}_1 = \partial^{}_w$, and the right-hand sides of the differential equations will be independent of $w$. In our case, the transformation of coordinates is trivial and we can simply identify $y = x^{}_-$, $u = x^{}_+$ and $w = a$. The differential equations are then converted into
\begin{eqnarray}
\frac{{\rm d}w}{{\rm d}y} &=& \frac{{\rm d}a}{{\rm d}x^{}_-} = -\frac{1}{x^{}_+ x^{}_-} \; , \label{eq:dwdy} \\
\frac{{\rm d}u}{{\rm d}y} &=& \frac{{\rm d}x^{}_+}{{\rm d}x^{}_-} = - \frac{x^2_- - 2 x^2_+}{x^{}_+ x^{}_-} \label{eq:dudy} \; ,
\end{eqnarray}
where $x^{}_+$ should be regarded as a function of $x^{}_-$. Thus Eq.~(\ref{eq:dwdy}) can be readily solved by the following quadrature
\begin{eqnarray}
a = -\int \frac{{\rm d}x^{}_-}{x^{}_+ x^{}_-} \; ,
\label{eq:a}
\end{eqnarray}
where the invariance of the differential equations under the translations $a \mapsto a + \varepsilon$ can now be understood as the freedom in choosing the integration constant. The next step is to find the solution to Eq.~(\ref{eq:dudy}) by applying the remaining symmetry ${\bf v}^{}_2 = -a\partial^{}_a + x^{}_+ \partial^{}_+ + x^{}_-\partial^{}_-$. Since Eq.~(\ref{eq:dudy}) is independent of $a$, the relevant vector field is actually ${\bf v}^{}_2 = x^{}_+ \partial^{}_+ + x^{}_-\partial^{}_-$. One has to look for another set of coordinates $z = z(x^{}_+, x^{}_-)$ and $w = w(x^{}_+, x^{}_-)$ such that ${\bf v}^{}_2(z) = 0$ and ${\bf v}^{}_2(w) = 1$. It is straightforward to see that $z = x^{}_+/x^{}_-$ and $w = \ln(x^{}_-)$ fulfill these requirements. Consequently, Eq.~(\ref{eq:dudy}) can be rewritten as
\begin{eqnarray}
\frac{{\rm d}w}{{\rm d}z} = \frac{{\rm d}w/{\rm d}x^{}_-}{{\rm d}z/{\rm d}x^{}_-} = \frac{x^{}_+ x^{}_-}{x^2_+ - x^2_-} = \frac{z}{z^2 - 1} \; , \label{eq:dwdz}
\end{eqnarray}
which can be integrated immediately
\begin{eqnarray}
w = \int \frac{z{\rm d}z}{z^2 - 1} = \frac{1}{2} \ln \left(1 - z^2\right) + c^\prime \; ,
\label{eq:w}
\end{eqnarray}
with $c^\prime$ being the integration constant and $|z| < 1$. Inserting $w = \ln(x^{}_-)$ and $z = x^{}_+/x^{}_-$ back into Eq.~(\ref{eq:w}), we arrive at
\begin{eqnarray}
\frac{x^2_- - x^2_+}{x^4_-} = c \; ,
\label{eq:invariant1}
\end{eqnarray}
where the constant $c = \exp(-2c^\prime)$ can be fixed by the initial condition at $a = 0$. Given the definitions of $x^{}_+$ and $x^{}_-$ in Eq.~(\ref{eq:xpm}), the identity in Eq.~(\ref{eq:invariant1}) turns out to be
\begin{eqnarray}
4|V^{}_{e1}|^2 |V^{}_{e2}|^2 \widetilde{\Delta}^2_{21} = c = 4|U^{}_{e1}|^2 |U^{}_{e2}|^2 \Delta^2_{21} \; ,
\label{eq:2nuid}
\end{eqnarray}
which is the desired relationship between the vacuum and matter mixing parameters. As we have been successful in solving $x^{}_+$ as a function of $x^{}_-$ in Eq.~(\ref{eq:invariant1}), namely,
\begin{eqnarray}
x^{}_+ = \sqrt{x^2_- - c x^4_-} \; ,
\label{eq:xpsol}
\end{eqnarray}
it should be plugged into Eq.~(\ref{eq:a}) in order to finally find $x^{}_-$ out. More explicitly, we have
\begin{eqnarray}
a = - \int \frac{{\rm d}x^{}_-}{x^{}_- \sqrt{x^2_- - c x^4_-}} = \sqrt{x^{-2}_- - c} + c^{\prime\prime} \; ,
\label{eq:newa}
\end{eqnarray}
where the integration constant can be determined by setting $a = 0$, i.e., $c^{\prime\prime} = - \sqrt{\Delta^2_{21} - c}$. Then Eq.~(\ref{eq:newa}) leads to the following new identity
\begin{eqnarray}
\sqrt{\widetilde{\Delta}^2_{21} -c} = a + \sqrt{\Delta^2_{21} -c}
\label{eq:2nuid21}
\end{eqnarray}
or equivalently
\begin{eqnarray}
\left(|V^{}_{e2}|^2 - |V^{}_{e1}|^2\right) \widetilde{\Delta}^{}_{21} = \left(|U^{}_{e2}|^2 - |U^{}_{e1}|^2\right) \Delta^{}_{21} + a \; ,
\label{eq:2nuid22}
\end{eqnarray}
where the explicit expressions of $c$ in Eq.~(\ref{eq:2nuid}) have been implemented.
\end{itemize}

Although those two identities in Eqs.~(\ref{eq:2nuid}) and (\ref{eq:2nuid22}) for two-flavor neutrino oscillations in matter are well known, we have reproduced them by explicitly solving the differential equations. From such a simple exercise, one can learn that the continuous symmetries of one-parameter Lie groups for the differential equations play a critically important role in finding the solutions. Furthermore, we investigate the differential invariants similar to those in Eqs.~(\ref{eq:Naumov}) and (\ref{eq:Vei}) but in the two-flavor case. Suppose ${\cal I}(a, x^{}_+, x^{}_-)$ to be such a kind of differential invariant, which is vanishing under the action of two vector fields ${\bf v}^{}_1 = \partial^{}_a$ and ${\bf w} = ({\rm d}x^{}_+/{\rm d}a) \partial^{}_+ + ({\rm d}x^{}_-/{\rm d}a) \partial^{}_-$. Notice that ${\bf v}^{}_1[{\cal I}(a, x^{}_+, x^{}_-)] = 0$ implies no explicit dependence on $a$, namely, ${\cal I}(a, x^{}_+, x^{}_-) = {\cal I}(x^{}_+, x^{}_-)$. On the other hand, ${\bf w}[{\cal I}(x^{}_+, x^{}_-)] = 0$ leads to the partial differential equation below
\begin{eqnarray}
\frac{{\rm d}x^{}_+}{{\rm d}a} \frac{\partial}{\partial x^{}_+} {\cal I}(x^{}_+, x^{}_-) + \frac{{\rm d}x^{}_-}{{\rm d}a} \frac{\partial}{\partial x^{}_-} {\cal I}(x^{}_+, x^{}_-) = 0 \; ,
\label{eq:partial}
\end{eqnarray}
whose characteristic equation gives rise to exactly the same result of ${\rm d}x^{}_+/{\rm d}x^{}_-$ in Eq.~(\ref{eq:dudy}). The solution to this equation has already been given in Eq.~(\ref{eq:invariant1}). According to {\bf Theorem 2.17} in Ref.~\cite{Olver}, the three-dimensional manifold $\mathbb{R}^3$ with the local coordinates $(a, x^{}_+, x^{}_-)$ is invariant under the Lie group $G$ with a two-dimensional orbit, for which the Lie algebra is $[{\bf v}^{}_1, {\bf w}] = 0$, so the group invariant ${\cal I}(x^{}_+, x^{}_-) = (x^2_- - x^2_+)/x^4_- = 4 |V^{}_{e1}|^2 |V^{}_{e2}|^2 \widetilde{\Delta}^2_{21}$ is unique.

\section{Three-flavor Neutrino Oscillations}
\label{sec:three}

We have demonstrated that the RGEs for two-flavor neutrino oscillations in matter can be solved by quadratures alone, as a consequence of the continuous symmetries corresponding to two one-parameter Lie groups. In this section, we turn to the case of three-flavor neutrino oscillations, but focus only on the differential equations in Eqs.~(\ref{eq:RGEe1})-(\ref{eq:Del31}) that are already in a closed form. The complete set of RGEs have been presented in Ref.~\cite{Xing:2018lob}, where the differential equations for the moduli of mixing matrix elements $|V^{}_{\alpha i}|^2$ (for $\alpha = e, \mu, \tau$ and $i = 1, 2, 3$) can be found. The symmetry analysis of this complete set of RGEs will be more involved.

\subsection{New Variables}

In the three-flavor case, one can always choose four independent variables, e.g., $\{|V^{}_{e1}|^2, |V^{}_{e2}|^2\}$ and $\{\widetilde{\Delta}^{}_{21}, \widetilde{\Delta}^{}_{31}\}$, but the $S^{}_3$ symmetry of the system will be broken. Inspired by the solutions in the two-flavor case, we should introduce new variables in order to rewrite Eqs.~(\ref{eq:RGEe1})-(\ref{eq:Del31}) but still in an invariant form under the $S^{}_3$ transformations. First of all, we define another set of six dependent variables, namely,
\begin{eqnarray}
&~& x^{}_1 \equiv \frac{|V^{}_{e2}|^2}{\widetilde{\Delta}^{}_{23}} \; , \quad y^{}_1 \equiv \frac{|V^{}_{e3}|^2}{\widetilde{\Delta}^{}_{23}} \; , \label{eq:x1y1}\\
&~& x^{}_2 \equiv \frac{|V^{}_{e3}|^2}{\widetilde{\Delta}^{}_{31}} \; , \quad y^{}_2 \equiv \frac{|V^{}_{e1}|^2}{\widetilde{\Delta}^{}_{31}} \; , \label{eq:x2y2}\\
&~& x^{}_3 \equiv \frac{|V^{}_{e1}|^2}{\widetilde{\Delta}^{}_{12}} \; , \quad y^{}_3 \equiv \frac{|V^{}_{e2}|^2}{\widetilde{\Delta}^{}_{12}} \; ,
\label{eq:x3y3}
\end{eqnarray}
which have to fulfill the following constraint conditions
\begin{eqnarray}
x^{}_1 x^{}_2 + x^{}_2 y^{}_3 + y^{}_3 y^{}_1 &=& 0 \; , \label{eq:constr1} \\
x^{}_2 x^{}_3 + x^{}_3 y^{}_1 + y^{}_1 y^{}_2 &=& 0 \; , \label{eq:constr2} \\
x^{}_3 x^{}_1 + x^{}_1 y^{}_2 + y^{}_2 y^{}_3 &=& 0 \; . \label{eq:constr3}
\end{eqnarray}
Note that only two of the above three constraints are independent. For example, if we multiply Eq.~(\ref{eq:constr1}) by $x^{}_3$ and subtract Eq.~(\ref{eq:constr2}) multiplied by $y^{}_3$ from it, then we can obtain $x^{}_1 x^{}_2 x^{}_3 = y^{}_1 y^{}_2 y^{}_3$. Moreover, Eq.~(\ref{eq:constr3}) can be derived by multiplying Eq.~(\ref{eq:constr2}) by $x^{}_1$ and then replacing $x^{}_1 x^{}_2 x^{}_3$ by $y^{}_1 y^{}_2 y^{}_3$. In view of the $S^{}_3$ symmetry, one may take two constraints as $x^{}_1 x^{}_2 x^{}_3 = y^{}_1 y^{}_2 y^{}_3$ and
\begin{eqnarray}
(x^{}_1 x^{}_2 + x^{}_2 x^{}_3 + x^{}_3 x^{}_1) + (y^{}_1 y^{}_2 + y^{}_2 y^{}_3 + y^{}_3 y^{}_1) + (x^{}_1 y^{}_2 + x^{}_2 y^{}_3 + x^{}_3 y^{}_1) = 0 \; .
\label{eq:constr4}
\end{eqnarray}
One can verify that under the permutations of three neutrino mass eigenstates, the newly defined variables transform as follows
\begin{eqnarray}
I: &~& x^{}_i \to x^{}_i \; , \quad y^{}_i \to y^{}_i \; , \\
S^{(12)}_{}: &~& x^{}_1 \leftrightarrow - y^{}_2 \; , \quad x^{}_2 \leftrightarrow - y^{}_1 \; , \quad x^{}_3 \leftrightarrow -y^{}_3 \; , \\
S^{(23)}_{}: &~& x^{}_1 \leftrightarrow - y^{}_1 \; , \quad x^{}_2 \leftrightarrow - y^{}_3 \; , \quad x^{}_3 \leftrightarrow -y^{}_2 \; , \\
S^{(31)}_{}: &~& x^{}_1 \leftrightarrow - y^{}_3 \; , \quad x^{}_2 \leftrightarrow - y^{}_2 \; , \quad x^{}_3 \leftrightarrow -y^{}_1 \; , \\
S^{(123)}_{}: &~& x^{}_1 \rightarrow x^{}_2 \rightarrow x^{}_3 \rightarrow x^{}_1 \; , \quad y^{}_1 \rightarrow y^{}_2 \rightarrow y^{}_3 \rightarrow y^{}_1 \; , \\
S^{(321)}_{}: &~& x^{}_1 \rightarrow x^{}_3 \rightarrow x^{}_2 \rightarrow x^{}_1 \; , \quad y^{}_1 \rightarrow y^{}_3 \rightarrow y^{}_2 \rightarrow y^{}_1 \; .
\end{eqnarray}
The basic properties of the $S^{}_3$ symmetry group with six elements $\{I, S^{(12)}_{}, S^{(23)}_{}, S^{(31)}_{}, S^{(123)}, S^{(321)}\}$ have been collected in Appendix~\ref{sec:app}. As the transformations under $\{S^{(12)}_{}, S^{(23)}_{}, S^{(31)}_{}\}$ mix up $x^{}_i$ and $y^{}_i$ (for $i = 1, 2, 3$), we shall invent another set of variables $z^{}_{+i} \equiv (x^{}_i - y^{}_i)/\sqrt{2}$ and $z^{}_{-i} \equiv (x^{}_i + y^{}_i)/\sqrt{2}$ (for $i = 1, 2, 3$), i.e.,
\begin{eqnarray}
z^{}_{+1} &=& \frac{|V^{}_{e2}|^2 - |V^{}_{e3}|^2}{\sqrt{2} \widetilde{\Delta}^{}_{23}} \; , \quad z^{}_{-1} = \frac{|V^{}_{e2}|^2 + |V^{}_{e3}|^2}{\sqrt{2} \widetilde{\Delta}^{}_{23}} \; ,
\label{eq:zp1zm1}\\
z^{}_{+2} &=& \frac{|V^{}_{e3}|^2 - |V^{}_{e1}|^2}{\sqrt{2} \widetilde{\Delta}^{}_{31}} \; , \quad z^{}_{-2} = \frac{|V^{}_{e3}|^2 + |V^{}_{e1}|^2}{\sqrt{2} \widetilde{\Delta}^{}_{31}} \; ,
\label{eq:zp2zm2}\\
z^{}_{+3} &=& \frac{|V^{}_{e1}|^2 - |V^{}_{e2}|^2}{\sqrt{2} \widetilde{\Delta}^{}_{12}} \; , \quad z^{}_{-3} = \frac{|V^{}_{e1}|^2 + |V^{}_{e2}|^2}{\sqrt{2} \widetilde{\Delta}^{}_{12}} \; .
\label{eq:zp3zm3}
\end{eqnarray}
Consequently, the $S^{}_3$ transformations act on $z^{}_{+i}$ and $z^{}_{-i}$ (for $i = 1, 2, 3$) separately
\begin{eqnarray}
I: &~& z^{}_{\pm i} \to z^{}_{\pm i} \; ,  \\
S^{(12)}_{}: &~& z^{}_{\pm 1} \leftrightarrow \pm z^{}_{\pm 2} \; , \quad z^{}_{\pm 3} \rightarrow \pm z^{}_{\pm 3} \; , \\
S^{(23)}_{}: &~& z^{}_{\pm 2} \leftrightarrow \pm z^{}_{\pm 3} \; , \quad z^{}_{\pm 1} \rightarrow \pm z^{}_{\pm 1} \; , \\
S^{(31)}_{}: &~& z^{}_{\pm 1} \leftrightarrow \pm z^{}_{\pm 3} \; , \quad z^{}_{\pm 2} \rightarrow \pm z^{}_{\pm 2} \; , \\
S^{(123)}_{}: &~& z^{}_{\pm 1} \rightarrow z^{}_{\pm 2} \rightarrow z^{}_{\pm 3} \rightarrow z^{}_{\pm 1} \; , \\
S^{(321)}_{}: &~& z^{}_{\pm 1} \rightarrow z^{}_{\pm 3} \rightarrow z^{}_{\pm 2} \rightarrow z^{}_{\pm 1} \; .
\end{eqnarray}
It is now evident that $z^{}_{+i}$ and $z^{}_{-i}$ are disentangled, and thus there is no mixing between these two sectors under the $S^{}_3$ symmetry transformations.

Then, we come to the RGEs for three-flavor neutrino oscillations in matter, but rewrite them in terms of $\{x^{}_i, y^{}_i\}$ or $\{z^{}_{+ i}, z^{}_{-i}\}$. Although only the latter will be used in our later discussions, we write down both of them for comparison. The RGEs for $\{x^{}_i, y^{}_i\}$ can be derived with the help of their definitions in Eqs.~(\ref{eq:x1y1})-(\ref{eq:x3y3}) and the RGEs for the effective oscillation parameters in Eqs.~(\ref{eq:RGEe1})-(\ref{eq:Del31}). However, because of the $S^{}_3$ symmetry, it is only necessary to find out one equation and then get the others by performing the $S^{}_3$ transformations. For instance, we derive the RGE for $x^{}_1$ first, namely,
\begin{eqnarray}
\frac{{\rm d}}{{\rm d}a} \left(\frac{|V^{}_{e2}|^2}{\widetilde{\Delta}^{}_{23}}\right) &=& \frac{1}{\widetilde{\Delta}^{}_{23}} \left(\frac{\rm d}{{\rm d}a} |V^{}_{e2}|^2\right) - \frac{|V^{}_{e2}|^2}{\widetilde{\Delta}^2_{23}}  \left(\frac{\rm d}{{\rm d}a} \widetilde{\Delta}^{}_{23}\right) \nonumber \\
&=& 2 \frac{|V^{}_{e2}|^2}{\widetilde{\Delta}^{}_{23}} \cdot \frac{|V^{}_{e3}|^2}{\widetilde{\Delta}^{}_{23}} - 2 \frac{|V^{}_{e2}|^2}{\widetilde{\Delta}^{}_{23}} \cdot \frac{|V^{}_{e1}|^2 }{\widetilde{\Delta}^{}_{12}} - \frac{|V^{}_{e2}|^2 }{\widetilde{\Delta}^{}_{23}} \left( \frac{|V^{}_{e2}|^2}{\widetilde{\Delta}^{}_{23}} - \frac{|V^{}_{e3}|^2 }{\widetilde{\Delta}^{}_{23}}\right) \; ,
\label{eq:dx1da}
\end{eqnarray}
which can be written as
\begin{eqnarray}
\frac{{\rm d}x^{}_1}{{\rm d}a} = 2 x^{}_1 y^{}_1 - 2 x^{}_1 x^{}_3 - x^{}_1 (x^{}_1 - y^{}_1) \; .
\label{eq:dx1da1}
\end{eqnarray}
The complete set of RGEs for $\{x^{}_i, y^{}_i\}$ can then be obtained from Eq.~(\ref{eq:dx1da1}) by applying the $S^{}_3$ transformations corresponding to all six elements $\{I, S^{(12)}_{}, S^{(23)}_{}, S^{(31)}_{}, S^{(123)}, S^{(321)}\}$. Though there are several different ways to apply these transformations, the final results should be the same. More explicitly, we apply the $S^{}_3$ transformations to Eq.~(\ref{eq:dx1da1}) and then arrive at
\begin{eqnarray}
I: &~& \frac{{\rm d}x^{}_1}{{\rm d}a} = 2 x^{}_1 y^{}_1 - 2 x^{}_1 x^{}_3 - x^{}_1 (x^{}_1 - y^{}_1) \; , \\
S^{(12)}_{}: &~& \frac{{\rm d}y^{}_2}{{\rm d}a} = 2 y^{}_2 y^{}_3 - 2 y^{}_2 x^{}_2 - y^{}_2 (x^{}_2 - y^{}_2) \; , \\
S^{(23)}_{}: &~& \frac{{\rm d}y^{}_1}{{\rm d}a} = 2 y^{}_1 y^{}_2 - 2 y^{}_1 x^{}_1 - y^{}_1 (x^{}_1 - y^{}_1) \; , \\
S^{(31)}_{}: &~& \frac{{\rm d}y^{}_3}{{\rm d}a} = 2 y^{}_3 y^{}_1 - 2 y^{}_3 x^{}_3 - y^{}_3 (x^{}_3 - y^{}_3) \; ,\\
S^{(123)}_{}: &~& \frac{{\rm d}x^{}_2}{{\rm d}a} = 2 x^{}_2 y^{}_2 - 2 x^{}_2 x^{}_1 - x^{}_2 (x^{}_2 - y^{}_2) \; , \\
S^{(321)}_{}: &~& \frac{{\rm d}x^{}_3}{{\rm d}a} = 2 x^{}_3 y^{}_3 - 2 x^{}_3 x^{}_2 - x^{}_3 (x^{}_3 - y^{}_3) \; .
\end{eqnarray}
In terms of $z^{}_{\pm i}$ for $i = 1, 2, 3$, the above equations can be rewritten as
\begin{eqnarray}
\frac{{\rm d}z^{}_{+1}}{{\rm d}a} &=& \sqrt{2} \left\{ z^2_{-1} - 2 z^2_{+1} - \frac{1}{2} \left[(z^{}_{+1} - z^{}_{-1})(z^{}_{+2} - z^{}_{-2}) + (z^{}_{+3} + z^{}_{-3})(z^{}_{+1} + z^{}_{-1})\right] \right\}\; , \label{eq:dzp1da}\\
\frac{{\rm d}z^{}_{+2}}{{\rm d}a} &=& \sqrt{2} \left\{ z^2_{-2} - 2 z^2_{+2} - \frac{1}{2} \left[(z^{}_{+2} - z^{}_{-2})(z^{}_{+3} - z^{}_{-3}) + (z^{}_{+1} + z^{}_{-1})(z^{}_{+2} + z^{}_{-2})\right] \right\}\; , \label{eq:dzp2da}\\
\frac{{\rm d}z^{}_{+3}}{{\rm d}a} &=& \sqrt{2} \left\{ z^2_{-3} - 2 z^2_{+3} - \frac{1}{2} \left[(z^{}_{+3} - z^{}_{-3})(z^{}_{+1} - z^{}_{-1}) + (z^{}_{+2} + z^{}_{-2})(z^{}_{+3} + z^{}_{-3})\right] \right\} \; , \label{eq:dzp3da}\\
\frac{{\rm d}z^{}_{-1}}{{\rm d}a} &=& - \sqrt{2} z^{}_{+1} z^{}_{-1} - \frac{1}{\sqrt{2}} \left[(z^{}_{+3} + z^{}_{-3})(z^{}_{+1} + z^{}_{-1}) - (z^{}_{+1} - z^{}_{-1})(z^{}_{+2} - z^{}_{-2})\right] \; , \label{eq:dzm1da}\\
\frac{{\rm d}z^{}_{-2}}{{\rm d}a} &=& - \sqrt{2} z^{}_{+2} z^{}_{-2} - \frac{1}{\sqrt{2}} \left[(z^{}_{+1} + z^{}_{-1})(z^{}_{+2} + z^{}_{-2}) - (z^{}_{+2} - z^{}_{-2})(z^{}_{+3} - z^{}_{-3})\right] \; , \label{eq:dzm2da} \\
\frac{{\rm d}z^{}_{-3}}{{\rm d}a} &=& - \sqrt{2} z^{}_{+3} z^{}_{-3} - \frac{1}{\sqrt{2}} \left[(z^{}_{+2} + z^{}_{-2})(z^{}_{+3} + z^{}_{-3}) - (z^{}_{+3} - z^{}_{-3})(z^{}_{+1} - z^{}_{-1})\right] \; . \label{eq:dzm3da}
\end{eqnarray}
It is worth mentioning that ${\bf z}^{}_+ \equiv (z^{}_{+1}, z^{}_{+2}, z^{}_{+3})^{\rm T}$ transforms as the reducible three-dimensional representation ${\bf 3}$ under the $S^{}_3$ symmetry, while ${\bf z}^{}_- \equiv (z^{}_{-1}, z^{}_{-2}, z^{}_{-3})^{\rm T}$ as another reducible three-dimensional representation ${\bf 3}^\prime$. The reduction of these two three-dimensional representations into irreducible representations of $S^{}_3$ is ${\bf 3} = {\bf 2} \oplus {\bf 1}$ and ${\bf 3}^\prime = {\bf 2} \oplus {\bf 1}^\prime$, respectively, where ${\bf 1}^\prime$ is the nontrivial one-dimensional representation.

\subsection{Irreducible Representations}

In the case of two-flavor neutrino oscillations, we have seen that $x^{}_+$ and $x^{}_-$ serve as the trivial and nontrivial one-dimensional representations of the $Z^{}_2$ group, respectively. In order to further simplify the coupled RGEs in Eqs.~(\ref{eq:dzp1da})-(\ref{eq:dzm3da}) for $z^{}_{\pm i}$, we shall convert them into the irreducible representations of the $S^{}_3$ group, for which three irreducible representations ${\bf 2}$, ${\bf 1}^\prime$ and ${\bf 1}$ exist. For this purpose, one can utilize two real orthogonal matrices
\begin{eqnarray}
R^{}_+ = \left(\begin{matrix} \displaystyle \frac{1}{\sqrt{2}} & \displaystyle -\frac{1}{\sqrt{2}} & 0 \cr \displaystyle \frac{1}{\sqrt{6}} & \displaystyle \frac{1}{\sqrt{6}} & \displaystyle -\frac{2}{\sqrt{6}} \cr \displaystyle \frac{1}{\sqrt{3}} & \displaystyle \frac{1}{\sqrt{3}} & \displaystyle \frac{1}{\sqrt{3}}\end{matrix}\right) \; , \quad R^{}_- = \left(\begin{matrix} \displaystyle \frac{1}{\sqrt{6}} & \displaystyle \frac{1}{\sqrt{6}} & \displaystyle -\frac{2}{\sqrt{6}} \cr \displaystyle -\frac{1}{\sqrt{2}} & \displaystyle \frac{1}{\sqrt{2}} & 0 \cr \displaystyle \frac{1}{\sqrt{3}} & \displaystyle \frac{1}{\sqrt{3}} & \displaystyle \frac{1}{\sqrt{3}}\end{matrix}\right) \; ,
\label{eq:RpRm}
\end{eqnarray}
to define two three-dimensional representations $\widetilde{\bf z}^{}_+ \equiv (\widetilde{z}^{}_{+1}, \widetilde{z}^{}_{+2}, \widetilde{z}^{}_{+3})^{\rm T}$ and $\widetilde{\bf z}^{}_- \equiv (\widetilde{z}^{}_{-1}, \widetilde{z}^{}_{-2}, \widetilde{z}^{}_{-3})^{\rm T}$, namely,
\begin{eqnarray}
\widetilde{\bf z}^{}_+ &=& R^{}_+ {\bf z}^{}_+ = \left(\begin{matrix} \displaystyle \frac{1}{\sqrt{2}} (z^{}_{+1} - z^{}_{+2}) \cr \displaystyle \frac{1}{\sqrt{6}} (z^{}_{+1} + z^{}_{+2} - 2z^{}_{+3}) \cr \displaystyle \frac{1}{\sqrt{3}} (z^{}_{+1} + z^{}_{+2} + z^{}_{+3}) \end{matrix}\right)\; , \label{eq:dztpda}\\
\widetilde{\bf z}^{}_- &=& R^{}_- {\bf z}^{}_- = \left(\begin{matrix} \displaystyle \frac{1}{\sqrt{6}} (z^{}_{-1} + z^{}_{-2} - 2z^{}_{-3}) \cr \displaystyle \frac{1}{\sqrt{2}} (z^{}_{-2} - z^{}_{-1}) \cr \displaystyle \frac{1}{\sqrt{3}} (z^{}_{-1} + z^{}_{-2} + z^{}_{-3}) \end{matrix}\right)\; , \label{eq:dztmda}
\end{eqnarray}
which have been decomposed into two irreducible representations as $\widetilde{\bf z}^{}_+ \sim {\bf 3} = \left[(\widetilde{z}^{}_{+1}, \widetilde{z}^{}_{+2})^{\rm T} \sim {\bf 2}\right] \oplus \left[\widetilde{z}^{}_{+3} \sim {\bf 1}\right]$ and $\widetilde{\bf z}^{}_- \sim {\bf 3}^\prime = \left[ (\widetilde{z}^{}_{-1}, \widetilde{z}^{}_{-2})^{\rm T} \sim {\bf 2}\right] \oplus \left[\widetilde{z}^{}_{-3} \sim {\bf 1}^\prime\right]$. The representation matrices of the $S^{}_3$ group elements and their reduced counterparts can be found in Appendix~\ref{sec:app}.

Making use of the RGEs of $z^{}_{\pm i}$ in Eqs.~(\ref{eq:dzp1da})-(\ref{eq:dzm3da}) and the relations between $z^{}_{\pm i}$ and $\widetilde{z}^{}_{\pm i}$ in Eqs.~(\ref{eq:dztpda})-(\ref{eq:dztmda}), one can get the RGEs of $\widetilde{z}^{}_{\pm i}$. Although the computations are straightforward, they are actually rather tedious. For conciseness, we omit all the calculational details and briefly present the basic properties that are useful for the practical calculations in Appendix~\ref{sec:app}. The final results are summarized as below
\begin{eqnarray}
\frac{{\rm d}}{{\rm d}a} D^{}_+ &=& \frac{1}{\sqrt{2}} D^{}_- S^{}_+ - \frac{3\sqrt{6}}{2} D^{}_+ S^{}_+ + \frac{\sqrt{6}}{2} \left(D^{}_- \otimes S^{}_-\right)^{}_{\bf 2} - \frac{1}{\sqrt{2}} \left(D^{}_+ \otimes S^{}_-\right)^{}_{\bf 2} \nonumber \\
&~& - \frac{\sqrt{3}}{2} \left(D^{}_+ \otimes D^{}_+\right)^{}_{\bf 2} - \frac{\sqrt{3}}{2} \left(D^{}_- \otimes D^{}_-\right)^{}_{\bf 2} + \left(D^{}_+ \otimes D^{}_-\right)^{}_{\bf 2}  \; , \label{eq:dDpda}\\
\frac{{\rm d}}{{\rm d}a} S^{}_+ &=& \frac{\sqrt{6}}{2} \left[\left(D^{}_- \otimes D^{}_-\right)^{}_{\bf 1} - \left(D^{}_+ \otimes D^{}_+\right)^{}_{\bf 1} - 2 S^2_+\right] \; , \label{eq:dSpda}\\
\frac{{\rm d}}{{\rm d}a} D^{}_- &=& -\frac{1}{\sqrt{2}} D^{}_+  S^{}_+ - \frac{\sqrt{6}}{2} D^{}_- S^{}_+ + \frac{\sqrt{6}}{2} \left(D^{}_+ \otimes S^{}_-\right)^{}_{\bf 2} -\frac{1}{\sqrt{2}} \left(D^{}_-\otimes S^{}_-\right)^{}_{\bf 2} \nonumber \\
&~&  + \frac{1}{2} \left(D^{}_+ \otimes D^{}_+\right)^{}_{\bf 2} - \frac{1}{2} \left(D^{}_- \otimes D^{}_-\right)^{}_{\bf 2} \; , \label{eq:dDmda}\\
\frac{{\rm d}}{{\rm d}a} S^{}_- &=& - \sqrt{6} S^{}_+ S^{}_- \; ,
\label{eq:dSmda}
\end{eqnarray}
where both $D^{}_+ \equiv (\widetilde{z}^{}_{+1}, \widetilde{z}^{}_{+2})^{\rm T} \sim {\bf 2}$ and $D^{}_- \equiv (\widetilde{z}^{}_{-1}, \widetilde{z}^{}_{-2})^{\rm T} \sim {\bf 2}$ are doublets of the $S^{}_3$ group, while $S^{}_+ \equiv \widetilde{z}^{}_{+3} \sim {\bf 1}$ and $S^{}_- \equiv \widetilde{z}^{}_{-3} \sim {\bf 1}^\prime$ are singlets. The rules for the decompositions of the direct products of two different irreducible representations have been given in the appendix as well, and the results relevant for Eqs.~(\ref{eq:dDpda})-(\ref{eq:dDmda}) are
\begin{eqnarray}
&~& \left(D^{}_+ \otimes S^{}_-\right)^{}_{\bf 2} = \left(\begin{matrix} -\widetilde{z}^{}_{+2} \widetilde{z}^{}_{-3} \cr + \widetilde{z}^{}_{+1} \widetilde{z}^{}_{-3} \end{matrix}\right) \; , \quad \left(D^{}_+ \otimes D^{}_+\right)^{}_{\bf 2} = \left(\begin{matrix} 2\widetilde{z}^{}_{+1} \widetilde{z}^{}_{+2} \cr  \widetilde{z}^2_{+1} - \widetilde{z}^2_{+2} \end{matrix}\right)\; , \\
&~& \left(D^{}_- \otimes S^{}_-\right)^{}_{\bf 2} = \left(\begin{matrix} -\widetilde{z}^{}_{-2} \widetilde{z}^{}_{-3} \cr + \widetilde{z}^{}_{-1} \widetilde{z}^{}_{-3} \end{matrix}\right) \; , \quad \left(D^{}_- \otimes D^{}_-\right)^{}_{\bf 2} = \left(\begin{matrix} 2\widetilde{z}^{}_{-1} \widetilde{z}^{}_{-2} \cr  \widetilde{z}^2_{-1} - \widetilde{z}^2_{-2} \end{matrix}\right) \; ,
\end{eqnarray}
and
\begin{eqnarray}
\left(D^{}_+ \otimes D^{}_-\right)^{}_{\bf 2} = \left(\begin{matrix} \widetilde{z}^{}_{+1} \widetilde{z}^{}_{-2} + \widetilde{z}^{}_{+2} \widetilde{z}^{}_{-1} \cr  \widetilde{z}^{}_{+1} \widetilde{z}^{}_{-1} - \widetilde{z}^{}_{+2} \widetilde{z}^{}_{-2} \end{matrix}\right) \; , \quad \left(D^{}_\pm \otimes D^{}_\pm\right)^{}_{\bf 1} = \widetilde{z}^2_{\pm 1} + \widetilde{z}^2_{\pm 2} \equiv D^2_\pm \; .
\end{eqnarray}
Some comments on the RGEs of $\widetilde{z}^{}_{\pm i}$ in Eqs.~(\ref{eq:dDpda})-(\ref{eq:dSmda}) are in order. First, these RGEs respect the $S^{}_3$ symmetry, under which ${\rm d}D^{}_\pm/{{\rm d}a}$ transform as doublets and ${\rm d}S^{}_\pm/{{\rm d}a}$ as singlets. Second, the RGEs of two singlets $S^{}_+$ and $S^{}_-$ are very simple. In particular, there is only one term on the right-hand side of Eq.~(\ref{eq:dSmda}). This simplicity can be well understood as follows. The right-hand side of Eq.~(\ref{eq:dSmda}) should be the polynomials in $\widetilde{z}^{}_{+ i}$ and $\widetilde{z}^{}_{-i}$ of order two, and must transform as the nontrivial singlet ${\bf 1}^\prime$, implying the unique choice of $S^{}_+ S^{}_-$. Third, on the right-hand side of Eq.~(\ref{eq:dSpda}), two extra singlets $D^2_+ \equiv \widetilde{z}^2_{+1} + \widetilde{z}^2_{+2}$ and $D^2_- \equiv \widetilde{z}^2_{-1} + \widetilde{z}^2_{-2}$ appear in addition to $S^2_+$. This observation is suggestive of looking for the RGEs of $D^2_+$ and $D^2_-$ instead. After some lengthy calculations, we obtain
\begin{eqnarray}
\frac{{\rm d}D^2_+}{{\rm d}a} &=& -3\sqrt{2} D^2_+ S^{}_+ + \sqrt{2} \left(D^{}_+ \otimes D^{}_-\right)^{}_{\bf 1} S^{}_+ + \sqrt{6} \left(D^{}_+ \otimes D^{}_-\right)^{}_{{\bf 1}^\prime} S^{}_- \nonumber \\
&~& + 2 \left(D^{}_+ \otimes D^{}_+ \otimes D^{}_-\right)^{}_{\bf 1} - \sqrt{3} \left(D^{}_+ \otimes D^{}_- \otimes D^{}_-\right)^{}_{\bf 1} -\sqrt{3} \left(D^{}_+ \otimes D^{}_+ \otimes D^{}_+\right)^{}_{\bf 1} \;, \label{eq:dDp2da}\\
\frac{{\rm d}D^2_-}{{\rm d}a} &=& -\sqrt{6} D^2_- S^{}_+ - \sqrt{2} \left(D^{}_+ \otimes D^{}_-\right)^{}_{\bf 1} S^{}_+ + \sqrt{6} \left(D^{}_+ \otimes D^{}_-\right)^{}_{{\bf 1}^\prime} S^{}_- \nonumber \\
&~& + \left(D^{}_+ \otimes D^{}_+ \otimes D^{}_-\right)^{}_{\bf 1} - \left(D^{}_- \otimes D^{}_- \otimes D^{}_-\right)^{}_{\bf 1} \; ,
\label{eq:dDm2da}
\end{eqnarray}
where the polynomials on the right-hand sides are of order three in $\widetilde{z}^{}_{\pm i}$ and the singlets constructed from two or three doublets can be figured out, i.e.,
\begin{eqnarray}
\left(D^{}_+ \otimes D^{}_-\right)^{}_{\bf 1} &=& \widetilde{z}^{}_{+1} \widetilde{z}^{}_{-1} + \widetilde{z}^{}_{+2} \widetilde{z}^{}_{-2} \; , \label{eq:news1}\\
\left(D^{}_+ \otimes D^{}_-\right)^{}_{{\bf 1}^\prime} &=& \widetilde{z}^{}_{+1} \widetilde{z}^{}_{-2} - \widetilde{z}^{}_{+2} \widetilde{z}^{}_{-1} \; ,
\label{eq:news2}\\
\left(D^{}_\pm \otimes D^{}_\pm \otimes D^{}_\mp\right)^{}_{\bf 1} &=& \left[ \left( \begin{matrix} 2 \widetilde{z}^{}_{\pm 1} \widetilde{z}^{}_{\pm 2} \cr \widetilde{z}^2_{\pm 1} - \widetilde{z}^2_{\pm 2}\end{matrix} \right) \otimes \left(\begin{matrix} \widetilde{z}^{}_{\mp 1} \cr \widetilde{z}^{}_{\mp 2}\end{matrix} \right) \right]^{}_{\bf 1} = 2 \widetilde{z}^{}_{\pm 1} \widetilde{z}^{}_{\pm 2} \widetilde{z}^{}_{\mp 1} + (\widetilde{z}^2_{\pm 1} - \widetilde{z}^2_{\pm 2}) \widetilde{z}^{}_{\mp 2} \; ,
\label{eq:news3}\\
\left(D^{}_\pm \otimes D^{}_\pm \otimes D^{}_\pm\right)^{}_{\bf 1} &=& \left[ \left( \begin{matrix} 2 \widetilde{z}^{}_{\pm 1} \widetilde{z}^{}_{\pm 2} \cr \widetilde{z}^2_{\pm 1} - \widetilde{z}^2_{\pm 2}\end{matrix} \right) \otimes \left(\begin{matrix} \widetilde{z}^{}_{\pm 1} \cr \widetilde{z}^{}_{\pm 2}\end{matrix} \right) \right]^{}_{\bf 1} = 3 \widetilde{z}^2_{\pm 1} \widetilde{z}^{}_{\pm 2} - \widetilde{z}^3_{\pm 2} \; .
\label{eq:news4}
\end{eqnarray}
It is easy to check that Eqs.~(\ref{eq:dDp2da}) and (\ref{eq:dDm2da}) are really singlets under the $S^{}_3$ symmetry. However, the presence of new singlets in Eqs.~(\ref{eq:news1})-(\ref{eq:news4}) renders it impossible to put the RGEs of $S^{}_\pm$ and $D^2_\pm$ in a closed form. Hence we cannot further simplify the RGEs in Eqs.~(\ref{eq:dDpda})-(\ref{eq:dSmda}).

\subsection{Continuous Symmetries}

So far we have implemented the $S^{}_3$ symmetry to simplify the RGEs as much as possible. Now we proceed to explore the continuous symmetries in this system of differential equations in Eqs.~(\ref{eq:dDpda})-(\ref{eq:dSmda}) following the same procedure as in the two-flavor case. In this subsection, we change the notations $D^{}_\pm$ and $S^{}_\pm$ back to the original variables $\widetilde{z}^{}_{\pm i}$, given the identification $D^{}_\pm \equiv (\widetilde{z}^{}_{\pm 1}, \widetilde{z}^{}_{\pm 2})^{\rm T}$ and $S^{}_\pm \equiv \widetilde{z}^{}_{\pm 3}$.

First, these differential equations are invariant under the one-parameter Lie group $(a, \widetilde{z}^{}_{\pm i}) \mapsto (a + \varepsilon, \widetilde{z}^{}_{\pm i})$ with $\varepsilon \in \mathbb{R}$ generated by the vector field ${\bf v}^{}_1 = \partial^{}_a$. As a result of this symmetry, we can reduce the total number of first-order differential equations by one, namely, Eq.~(\ref{eq:dSmda}) can be solved by a quadrature
\begin{eqnarray}
a = -\frac{1}{\sqrt{6}}\int \frac{{\rm d} \widetilde{z}^{}_{-3}}{ \widetilde{z}^{}_{+3} \widetilde{z}^{}_{-3}} \; ,
\label{eq:dadSm}
\end{eqnarray}
where $\widetilde{z}^{}_{+3}$ has been regarded as a function of $\widetilde{z}^{}_{-3}$.

Next, the scaling transformations $(a, \widetilde{z}^{}_{\pm i}) \mapsto (\lambda^{-1} a, \lambda \widetilde{z}^{}_{\pm i})$, where $\lambda$ is a nonzero real parameter, generated by another vector field ${\bf v}^{}_2 = -a\partial^{}_a + \widetilde{z}^{}_{\pm i} \partial^{}_{\pm i}$ with $\partial^{}_{\pm i} \equiv \partial/\partial \widetilde{z}^{}_{\pm i}$, leave the system of differential equations unchanged. After the quadrature in Eq.~(\ref{eq:dadSm}), the variable $a$ disappears from the system and the second vector field becomes ${\bf v}^{}_2 = \widetilde{z}^{}_{\pm i} \partial^{}_{\pm i}$, where the summation over the repeated index $i = 1, 2, 3$ is implied. This result can be ascribed to a suitable transformation of local coordinates, but the same notations of the coordinates are adopted. As we have seen in the two-flavor case, based on the scaling symmetry, we can define a new set of local coordinates
\begin{eqnarray}
y \equiv \ln \widetilde{z}^{}_{-3} \; , \quad w^{}_{+3} \equiv \frac{\widetilde{z}^{}_{+3}}{\widetilde{z}^{}_{-3}}  \;, \quad w^{}_{\pm k} \equiv \frac{\widetilde{z}^{}_{\pm k}}{\widetilde{z}^{}_{-3}} \; ,
\label{eq:wy}
\end{eqnarray}
with $k = 1, 2$. Two comments on these new coordinates are helpful. First, since $\widetilde{z}^{}_{+3}$ and $\widetilde{z}^{}_{-3}$ are singlets under the $S^{}_3$ symmetry, the above transformations indeed respect this discrete symmetry. One can observe that $(w^{}_{\pm 1}, w^{}_{\pm 2})^{\rm T}$ essentially transform as the $S^{}_3$ doublets. Second, it is straightforward to verify that ${\bf v}^{}_2(y) = 1$ and ${\bf v}^{}_2(w^{}_{\pm k}) = {\bf v}^{}_2(w^{}_{+3}) = 0$ for $k = 1, 2$. This property will help us to reduce the number of differential equations by one as well. More explicitly, we have
\begin{eqnarray}
\frac{{\rm d}y}{{\rm d}w^{}_{+3}} &=& \frac{{\rm d}y/{\rm d}\widetilde{z}^{}_{-3}}{{\rm d}w^{}_{+3}/{\rm d}\widetilde{z}^{}_{-3}} = \left(\frac{{\rm d}\widetilde{z}^{}_{+3}}{{\rm d}\widetilde{z}^{}_{-3}} - \frac{\widetilde{z}^{}_{+3}}{\widetilde{z}^{}_{-3}}\right)^{-1} \; ,
\label{eq:dydwp3}
\end{eqnarray}
where ${\rm d}\widetilde{z}^{}_{+3}/{\rm d}\widetilde{z}^{}_{-3}$ can be obtained from Eqs.~(\ref{eq:dSpda}) and (\ref{eq:dSmda}) as
\begin{eqnarray}
\frac{{\rm d}\widetilde{z}^{}_{+3}}{{\rm d}\widetilde{z}^{}_{-3}} = \frac{\widetilde{z}^{}_{+3}}{\widetilde{z}^{}_{-3}}  + \frac{1}{2} \left[\left( \frac{\widetilde{z}^{}_{+1}}{\widetilde{z}^{}_{-3}}\right)^2  \frac{\widetilde{z}^{}_{-3}}{\widetilde{z}^{}_{+3}} + \left( \frac{\widetilde{z}^{}_{+2}}{\widetilde{z}^{}_{-3}}\right)^2  \frac{\widetilde{z}^{}_{-3}}{\widetilde{z}^{}_{+3}} \right] - \frac{1}{2} \left[\left( \frac{\widetilde{z}^{}_{-1}}{\widetilde{z}^{}_{-3}}\right)^2  \frac{\widetilde{z}^{}_{-3}}{\widetilde{z}^{}_{+3}} + \left( \frac{\widetilde{z}^{}_{-2}}{\widetilde{z}^{}_{-3}}\right)^2  \frac{\widetilde{z}^{}_{-3}}{\widetilde{z}^{}_{+3}} \right] \; .
\label{eq:dzp3dzm3}
\end{eqnarray}
Combining Eqs.~(\ref{eq:dydwp3}) and (\ref{eq:dzp3dzm3}), we arrive at
\begin{eqnarray}
\frac{{\rm d}y}{{\rm d}w^{}_{+3}} &=& \frac{2 w^{}_{+3}}{(w^2_{+1} + w^2_{+2}) - (w^2_{-1} + w^2_{-2})} \; ,
\label{eq:dydwp3n}
\end{eqnarray}
which can be solved immediately
\begin{eqnarray}
y = \int \frac{2 w^{}_{+3} {\rm d}w^{}_{+3}}{(w^2_{+1} + w^2_{+2}) - (w^2_{-1} + w^2_{-2})} \; ,
\label{eq:ysol}
\end{eqnarray}
where $w^{}_{\pm k}$ for $k = 1, 2$ have been viewed as functions of $w^{}_{+3}$. Suppose that $w^{}_{\pm k}$ can be completely solved, then Eq.~(\ref{eq:ysol})
actually indicates that $\widetilde{z}^{}_{+3}$ (from $w^{}_{+3} = \widetilde{z}^{}_{+3}/\widetilde{z}^{}_{-3}$) is a function of $\widetilde{z}^{}_{-3}$ (from $y = \ln \widetilde{z}^{}_{-3}$) with the help of the inverse function theorem. Together with Eq.~(\ref{eq:dadSm}), this solution in Eq.~(\ref{eq:ysol}) completes the searches for $\widetilde{z}^{}_{+3}$ and $\widetilde{z}^{}_{-3}$ as functions of the matter parameter $a$. The remaining part of our task is then to find out $w^{}_{\pm k}$ as functions of $w^{}_{+3}$ from four differential equations of the $S^{}_3$ doublets, namely, Eqs.~(\ref{eq:dDpda}) and (\ref{eq:dDmda}).

Finally, we come to the most difficult part of the problem, since two continuous symmetries corresponding to two one-parameter Lie groups generated by ${\bf v}^{}_1$ and ${\bf v}^{}_2$ have been fully utilized. For the differential equations of the $S^{}_3$ doublets, we have
\begin{eqnarray}
\frac{{\rm d}w^{}_{\pm k}}{{\rm d}w^{}_{+3}} & = & \frac{{\rm d}w^{}_{\pm k}/{\rm d}\widetilde{z}^{}_{-3}}{{\rm d}w^{}_{+3}/{\rm d}\widetilde{z}^{}_{-3}} = \frac{({\rm d}\widetilde{z}^{}_{\pm k}/{\rm d}\widetilde{z}^{}_{-3}) - (\widetilde{z}^{}_{\pm k}/\widetilde{z}^{}_{-3})}{({\rm d}\widetilde{z}^{}_{+3}/{\rm d}\widetilde{z}^{}_{-3}) - (\widetilde{z}^{}_{+3}/\widetilde{z}^{}_{-3})} \; ,
\end{eqnarray}
whose explicit expressions can simply be derived using Eqs.~(\ref{eq:dDpda})-(\ref{eq:dSmda}), similar to that for ${\rm d}\widetilde{z}^{}_{+3}/{\rm d}\widetilde{z}^{}_{-3}$ in Eq.~(\ref{eq:dzp3dzm3}). After doing so, we find
\begin{eqnarray}
\frac{{\rm d}w^{}_{+1}}{{\rm d}w^{}_{+3}} &=& \frac{(\sqrt{3} w^{}_{+1} - w^{}_{-1})w^{}_{+3} - (w^{}_{+2} - \sqrt{3} w^{}_{-2})}{\sqrt{3} \left[(w^2_{+1} + w^2_{+2}) - (w^2_{-1} + w^2_{-2})\right]} + \frac{\sqrt{2}(w^{}_{+1} w^{}_{+2} + w^{}_{-1} w^{}_{-2})}{\left[(w^2_{+1} + w^2_{+2}) - (w^2_{-1} + w^2_{-2})\right]} \nonumber \\
&~& - \frac{2 (w^{}_{+1}w^{}_{-2} + w^{}_{+2} w^{}_{-1})}{\sqrt{6} \left[(w^2_{+1} + w^2_{+2}) - (w^2_{-1} + w^2_{-2})\right]} \; , \label{eq:dwp1dwp3} \\
\frac{{\rm d}w^{}_{+2}}{{\rm d}w^{}_{+3}} &=& \frac{(\sqrt{3} w^{}_{+2} - w^{}_{-2})w^{}_{+3} + (w^{}_{+1} - \sqrt{3} w^{}_{-1})}{\sqrt{3} \left[(w^2_{+1} + w^2_{+2}) - (w^2_{-1} + w^2_{-2})\right]} + \frac{(w^2_{+1} - w^2_{+2}) + (w^2_{-1} - w^2_{-2})}{\sqrt{2} \left[(w^2_{+1} + w^2_{+2}) - (w^2_{-1} + w^2_{-2})\right]} \nonumber \\
&~& - \frac{2 (w^{}_{+1}w^{}_{-1} - w^{}_{+2} w^{}_{-2})}{\sqrt{6} \left[(w^2_{+1} + w^2_{+2}) - (w^2_{-1} + w^2_{-2})\right]} \; , \label{eq:dwp2dwp3} \\
\frac{{\rm d}w^{}_{-1}}{{\rm d}w^{}_{+3}} &=& \frac{(w^{}_{+1} - \sqrt{3} w^{}_{-1})w^{}_{+3} + (\sqrt{3} w^{}_{+2} - w^{}_{-2})}{\sqrt{3} \left[(w^2_{+1} + w^2_{+2}) - (w^2_{-1} + w^2_{-2})\right]} - \frac{2(w^{}_{+1} w^{}_{+2} - w^{}_{-1} w^{}_{-2})}{\sqrt{6}\left[(w^2_{+1} + w^2_{+2}) - (w^2_{-1} + w^2_{-2})\right]} \;, \quad \label{eq:dwm1dwp3} \\
\frac{{\rm d}w^{}_{-2}}{{\rm d}w^{}_{+3}} &=& \frac{( w^{}_{+2} - \sqrt{3} w^{}_{-2})w^{}_{+3} - (\sqrt{3}w^{}_{+1} - w^{}_{-1})}{\sqrt{3} \left[(w^2_{+1} + w^2_{+2}) - (w^2_{-1} + w^2_{-2})\right]} - \frac{(w^2_{+1} - w^2_{+2}) - (w^2_{-1} - w^2_{-2})}{\sqrt{6}\left[(w^2_{+1} + w^2_{+2}) - (w^2_{-1} + w^2_{-2})\right]} \; .
\label{eq:dwm2dwp3}
\end{eqnarray}
Notice that $({\rm d}w^{}_{+1}/{\rm d}w^{}_{+3}, {\rm d}w^{}_{+2}/{\rm d}w^{}_{+3})^{\rm T}$ and $({\rm d}w^{}_{-1}/{\rm d}w^{}_{+3}, {\rm d}w^{}_{-2}/{\rm d}w^{}_{+3})^{\rm T}$ obviously transform as the irreducible representation ${\bf 2}$ of the $S^{}_3$ group. Unfortunately, one can check that the above set of differential equations are no longer invariant under the one-dimensional translations $(w^{}_{+3}, w^{}_{\pm k}) \mapsto (w^{}_{+3} + \varepsilon, w^{}_{\pm k})$ or the scaling transformations $(w^{}_{+3}, w^{}_{\pm k}) \mapsto (\lambda^{-1} w^{}_{+3}, \lambda w^{}_{\pm k})$ for $k = 1, 2$. Therefore, unlike the two-flavor case, the reduced RGEs in Eqs.~(\ref{eq:dwp1dwp3})-(\ref{eq:dwm2dwp3}) in the three-flavor case cannot be solved by quadratures alone.

As for the differential invariants ${\cal I}(a, \{\widetilde{z}^{}_{\pm i}\})$, we require ${\bf v}^{}_1\left[{\cal I}(a, \{\widetilde{z}^{}_{\pm i}\})\right] = \partial^{}_a {\cal I}(a, \{\widetilde{z}^{}_{\pm i}\})= 0$, namely, ${\cal I}(a, \{\widetilde{z}^{}_{\pm i}\}) = {\cal I}(\{\widetilde{z}^{}_{\pm i}\})$ does not explicitly depend on $a$. Moreover, ${\cal I}(\{\widetilde{z}^{}_{\pm i}\})$ should also be invariant functions of the vector field ${\bf w} = ({\rm d}\widetilde{z}^{}_{+i}/{\rm d}a) \partial^{}_{+i} + ({\rm d}\widetilde{z}^{}_{-i}/{\rm d}a) \partial^{}_{-i}$, i.e.,
\begin{eqnarray}
{\bf w} \left[{\cal I}(\{\widetilde{z}^{}_{\pm i}\})\right] = \sum^3_{j = 1} \left[ \frac{{\rm d}\widetilde{z}^{}_{+j}}{{\rm d}a} \cdot \frac{\partial {\cal I}(\{\widetilde{z}^{}_{\pm i}\})}{\partial \widetilde{z}^{}_{+j}} + \frac{{\rm d}\widetilde{z}^{}_{-j}}{{\rm d}a} \cdot \frac{\partial {\cal I}(\{\widetilde{z}^{}_{\pm i}\})}{\partial \widetilde{z}^{}_{-j}} \right] = 0 \; .
\label{eq:invariant2}
\end{eqnarray}
The characteristic equation of the above partial differential equation gives rise to five first-order ordinary differential equations, which are actually equivalent to those in Eqs.~(\ref{eq:dDpda})-(\ref{eq:dDmda}) but divided by Eq.~(\ref{eq:dSmda}) on both sides. Hence the search for the desired differential invariants is as difficult as solving the original first-order differential equations. According to {\bf Theorem 5.17} in Ref.~\cite{Olver}, for the seven-dimensional manifold with the local coordinates $(a, \{\widetilde{z}^{}_{\pm i}\})$ or the initial ones $(a, \{x^{}_i\}, \{y^{}_i\})$, the invariant group generated by $\{{\bf v}^{}_1, {\bf w}\}$ is two-dimensional, so we should have five functionally-independent invariants in total. All the other invariants of $\{{\bf v}^{}_1, {\bf w}\}$ can be expressed as smooth functions of five independent ones. However, as we have mentioned, there exist four constraint conditions, in which four functions $x^{}_1 x^{}_2 + x^{}_2 y^{}_3 + y^{}_3 y^{}_1$, $x^{}_2 x^{}_3 + x^{}_3 y^{}_1 + y^{}_1 y^{}_2$, $x^{}_3 x^{}_1 + x^{}_1 y^{}_2 + y^{}_2 y^{}_3$ from Eqs.~(\ref{eq:constr1})-(\ref{eq:constr3}) and $x^{}_1 x^{}_2 x^{}_3 - y^{}_1 y^{}_2 y^{}_3$ are involved. As functions, they are linearly independent, and all of them vanish when $x^{}_i$ and $y^{}_i$ are the solutions to the RGEs. Therefore, we can in principle solve $\{w^{}_{+1}, w^{}_{+2}, w^{}_{-1}, w^{}_{-2}\}$ as functions of $w^{}_{+3}$ from these four constraint conditions, and insert them into Eq.~(\ref{eq:ysol}), leading to a unique differential invariant in Eq.~(\ref{eq:Vei}). The explicit construction of this invariant is in fact complicated and will be left for a future work. In addition, the construction of the Naumov relation in Eq.~(\ref{eq:Naumov}) directly from the RGEs of all the elements of the effective mixing matrix $V^{}_{\alpha i}$ (for $\alpha = e, \mu, \tau$ and $i = 1, 2, 3$) and neutrino mass-squared differences $\widetilde{\Delta}^{}_{ij}$ (for $ij = 12, 23, 31$) demands even more efforts.

\section{Further Remarks}\label{sec:remark}

In this section, we give some brief remarks on the relations between the basic oscillation parameters in vacuum and the effective ones in matter, which should be associated with the solutions to the differential equations of our interest. As has been demonstrated in Ref.~\cite{Xing:2019owb}, one can establish the following identities
\begin{eqnarray}
	|V^{}_{e1}|^2 &=&  - \frac{ \left[ (\widetilde{\Delta}^{}_{31} - \widetilde{\Delta}^{}_{12}) - (\Delta^{}_{23} - \Delta^{}_{31}) - a\right] \cdot \left[ (\widetilde{\Delta}^{}_{31} - \widetilde{\Delta}^{}_{12}) - (\Delta^{}_{12} - \Delta^{}_{23}) - a \right]}{9 \widetilde{\Delta}^{}_{31} \widetilde{\Delta}^{}_{12}} |U^{}_{e1}|^2  \nonumber \\
	&~& - \frac{ \left[ (\widetilde{\Delta}^{}_{31} - \widetilde{\Delta}^{}_{12}) - (\Delta^{}_{23} - \Delta^{}_{31}) - a\right] \cdot \left[ (\widetilde{\Delta}^{}_{31} - \widetilde{\Delta}^{}_{12}) - (\Delta^{}_{31} - \Delta^{}_{12}) - a \right]}{9 \widetilde{\Delta}^{}_{31} \widetilde{\Delta}^{}_{12}} |U^{}_{e2}|^2 \nonumber \\
	&~& - \frac{ \left[ (\widetilde{\Delta}^{}_{31} - \widetilde{\Delta}^{}_{12}) - (\Delta^{}_{12} - \Delta^{}_{23}) - a\right] \cdot \left[ (\widetilde{\Delta}^{}_{31} - \widetilde{\Delta}^{}_{12}) - (\Delta^{}_{31} - \Delta^{}_{12}) - a \right]}{9 \widetilde{\Delta}^{}_{31} \widetilde{\Delta}^{}_{12}} |U^{}_{e3}|^2 \; , \label{eq:Ve1Uej} \quad \\
	|V^{}_{e2}|^2 &=&  - \frac{ \left[ (\widetilde{\Delta}^{}_{12} - \widetilde{\Delta}^{}_{23}) - (\Delta^{}_{23} - \Delta^{}_{31}) - a\right] \cdot \left[ (\widetilde{\Delta}^{}_{12} - \widetilde{\Delta}^{}_{23}) - (\Delta^{}_{12} - \Delta^{}_{23}) - a \right]}{9 \widetilde{\Delta}^{}_{12} \widetilde{\Delta}^{}_{23}} |U^{}_{e1}|^2  \nonumber \\
	&~& - \frac{ \left[ (\widetilde{\Delta}^{}_{12} - \widetilde{\Delta}^{}_{23}) - (\Delta^{}_{23} - \Delta^{}_{31}) - a\right] \cdot \left[ (\widetilde{\Delta}^{}_{12} - \widetilde{\Delta}^{}_{23}) - (\Delta^{}_{31} - \Delta^{}_{12}) - a \right]}{9 \widetilde{\Delta}^{}_{12} \widetilde{\Delta}^{}_{23}} |U^{}_{e2}|^2 \nonumber \\
	&~& - \frac{ \left[ (\widetilde{\Delta}^{}_{12} - \widetilde{\Delta}^{}_{23}) - (\Delta^{}_{12} - \Delta^{}_{23}) - a\right] \cdot \left[ (\widetilde{\Delta}^{}_{12} - \widetilde{\Delta}^{}_{23}) - (\Delta^{}_{31} - \Delta^{}_{12}) - a \right]}{9 \widetilde{\Delta}^{}_{12} \widetilde{\Delta}^{}_{23}} |U^{}_{e3}|^2 \; , \label{eq:Ve2Uej} \quad \\
	|V^{}_{e3}|^2 &=&  - \frac{ \left[ (\widetilde{\Delta}^{}_{23} - \widetilde{\Delta}^{}_{31}) - (\Delta^{}_{23} - \Delta^{}_{31}) - a\right] \cdot \left[ (\widetilde{\Delta}^{}_{23} - \widetilde{\Delta}^{}_{31}) - (\Delta^{}_{12} - \Delta^{}_{23}) - a \right]}{9 \widetilde{\Delta}^{}_{23} \widetilde{\Delta}^{}_{31}} |U^{}_{e1}|^2  \nonumber \\
    &~& - \frac{ \left[ (\widetilde{\Delta}^{}_{23} - \widetilde{\Delta}^{}_{31}) - (\Delta^{}_{23} - \Delta^{}_{31}) - a\right] \cdot \left[ (\widetilde{\Delta}^{}_{23} - \widetilde{\Delta}^{}_{31}) - (\Delta^{}_{31} - \Delta^{}_{12}) - a \right]}{9 \widetilde{\Delta}^{}_{23} \widetilde{\Delta}^{}_{31}} |U^{}_{e2}|^2 \nonumber \\
    &~& - \frac{ \left[ (\widetilde{\Delta}^{}_{23} - \widetilde{\Delta}^{}_{31}) - (\Delta^{}_{12} - \Delta^{}_{23}) - a\right] \cdot \left[ (\widetilde{\Delta}^{}_{23} - \widetilde{\Delta}^{}_{31}) - (\Delta^{}_{31} - \Delta^{}_{12}) - a \right]}{9 \widetilde{\Delta}^{}_{23} \widetilde{\Delta}^{}_{31}} |U^{}_{e3}|^2 \; , \label{eq:Ve3Uej} \quad
\end{eqnarray}
where all the neutrino mass-squared differences $\widetilde{\Delta}^{}_{ij}$ and $\Delta^{}_{ij}$ (for $ij = 12, 23, 31$) have been used to make the permutation symmetry $S^{}_3$ manifest while only the independent mass-squared differences $\{\widetilde{\Delta}^{}_{21}, \widetilde{\Delta}^{}_{31}\}$ and $\{\Delta^{}_{21}, \Delta^{}_{31}\}$ have been implemented in Ref.~\cite{Xing:2019owb}. Some comments on the identities in Eqs.~(\ref{eq:Ve1Uej})-(\ref{eq:Ve3Uej}) are in order. 

First, as extensively studied in Ref.~\cite{Denton:2019pka} and in earlier mathematical literature, the eigenvectors of the Hermitian matrix with distinct eigenvalues can be expressed in terms of its eigenvalues and the eigenvalues of its principal submatrices~\cite{Denton:2019ovn}. In the present case, the Hermitian matrix is just the effective Hamiltonian in Eq.~(\ref{eq:Hm}) for three-flavor neutrino oscillations in matter, so the eigenvector-eigenvalue theorem can also be applied to derive those identities in Eqs.~(\ref{eq:Ve1Uej})-(\ref{eq:Ve3Uej}). 

Second, in comparison with the expressions of $|V^{}_{ei}|^2$ (for $i = 1, 2, 3$) in Ref.~\cite{Xing:2019owb}, those in Eqs.~(\ref{eq:Ve1Uej})-(\ref{eq:Ve3Uej}) are obviously more symmetric, as we shall explain now. It is straightforward to verify that $(\widetilde{\Delta}^{}_{31} - \widetilde{\Delta}^{}_{12}, \widetilde{\Delta}^{}_{12} - \widetilde{\Delta}^{}_{23}, \widetilde{\Delta}^{}_{23} - \widetilde{\Delta}^{}_{31})$, $(\widetilde{\Delta}^{}_{31} \widetilde{\Delta}^{}_{12}, \widetilde{\Delta}^{}_{12} \widetilde{\Delta}^{}_{23}, \widetilde{\Delta}^{}_{23} \widetilde{\Delta}^{}_{31})$ and $(\Delta^{}_{31} - \Delta^{}_{12}, \Delta^{}_{12} - \Delta^{}_{23}, \Delta^{}_{23} - \Delta^{}_{31})$ transform as the reducible three-dimensional representation ${\bf 3}$ under the permutation group $S^{}_3$ acting on the indices $\{1, 2, 3\}$ for three neutrino mass eigenstates. In addition, both $(|V^{}_{e1}|^2, |V^{}_{e2}|^2, |V^{}_{e3}|^2)$ and $(|U^{}_{e1}|^2, |U^{}_{e2}|^2, |U^{}_{e3}|^2)$ transform exactly in the same way. With these transformation rules in mind, one can recast the identities in Eqs.~(\ref{eq:Ve1Uej})-(\ref{eq:Ve3Uej}) into
\begin{eqnarray}
	|V^{}_{e1}|^2 &=& - \frac{ \left[ (\widetilde{\Delta}^{}_{31} - \widetilde{\Delta}^{}_{12}) - a\right]^2 + \left[ (\widetilde{\Delta}^{}_{31} - \widetilde{\Delta}^{}_{12}) - a\right] S^{}_{\Delta U} + S^{}_{\Delta^2 U}}{9 \widetilde{\Delta}^{}_{31} \widetilde{\Delta}^{}_{12}} \; ,  \label{eq:Ve1Uejnew} \\
	|V^{}_{e2}|^2 &=&  - \frac{ \left[ (\widetilde{\Delta}^{}_{12} - \widetilde{\Delta}^{}_{23}) - a\right]^2 + \left[ (\widetilde{\Delta}^{}_{12} - \widetilde{\Delta}^{}_{23}) - a\right] S^{}_{\Delta U} + S^{}_{\Delta^2 U}}{9 \widetilde{\Delta}^{}_{12} \widetilde{\Delta}^{}_{23}} \; , \label{eq:Ve2Uejnew} \\
	|V^{}_{e3}|^2 &=&  - \frac{ \left[ (\widetilde{\Delta}^{}_{23} - \widetilde{\Delta}^{}_{31}) - a\right]^2 +  \left[ (\widetilde{\Delta}^{}_{23} - \widetilde{\Delta}^{}_{31}) - a\right] S^{}_{\Delta U} + S^{}_{\Delta^2 U}}{9 \widetilde{\Delta}^{}_{23} \widetilde{\Delta}^{}_{31}} \; , \label{eq:Ve3Uejnew} 
\end{eqnarray}
where it should be noticed that $(\Delta^{}_{31} - \Delta^{}_{12})|U^{}_{e1}|^2 + (\Delta^{}_{12} - \Delta^{}_{23})|U^{}_{e2}|^2 + (\Delta^{}_{23} - \Delta^{}_{31})|U^{}_{e3}|^2 \equiv S^{}_{\Delta U}$ and $(\Delta^{}_{12} - \Delta^{}_{23})(\Delta^{}_{23} - \Delta^{}_{31}) |U^{}_{e1}|^2 + (\Delta^{}_{23} - \Delta^{}_{31})(\Delta^{}_{31} - \Delta^{}_{12}) |U^{}_{e2}|^2 + (\Delta^{}_{31} - \Delta^{}_{12})(\Delta^{}_{12} - \Delta^{}_{23}) |U^{}_{e3}|^2 \equiv S^{}_{\Delta^2 U}$ are actually invariant under the permutations. Therefore, $(|V^{}_{e1}|^2, |V^{}_{e2}|^2, |V^{}_{e3}|^2)$ transforms in the same way as $(\widetilde{\Delta}^{}_{31} - \widetilde{\Delta}^{}_{12}, \widetilde{\Delta}^{}_{12} - \widetilde{\Delta}^{}_{23}, \widetilde{\Delta}^{}_{23} - \widetilde{\Delta}^{}_{31})$ and $(\widetilde{\Delta}^{}_{31} \widetilde{\Delta}^{}_{12}, \widetilde{\Delta}^{}_{12} \widetilde{\Delta}^{}_{23}, \widetilde{\Delta}^{}_{23} \widetilde{\Delta}^{}_{31})$ do.

Finally, we point out that the sum rules for $(|V^{}_{e1}|^2, |V^{}_{e2}|^2, |V^{}_{e3}|^2)$ given in Eqs.~(\ref{eq:Ve1Uej})-(\ref{eq:Ve3Uej}), or equivalently those in Eqs.~(\ref{eq:Ve1Uejnew})-(\ref{eq:Ve3Uejnew}), are not the explicit solutions to the differential equations, since $(\widetilde{\Delta}^{}_{12}, \widetilde{\Delta}^{}_{23}, \widetilde{\Delta}^{}_{31})$ have not been explicitly expressed as functions of the matter parameter $a$. However, since the eigenvalues $\{\widetilde{m}^2_1/(2E), \widetilde{m}^2_2/(2E), \widetilde{m}^2_3/(2E)\}$ of $H^{}_{\rm m}$ in Eq.~(\ref{eq:Hm}) can be obtained by solving the characteristic equation of $H^{}_{\rm m}$, we can indeed calculate $\widetilde{\Delta}^{}_{ij} = \widetilde{m}^2_i - \widetilde{m}^2_j$ (for $ij = 12, 23, 31$) as functions of $a$. Although the ultimate formulas of $|V^{}_{ei}|^2$ (for $i = 1, 2, 3$) are complicated, the results in Eqs.~(\ref{eq:Ve1Uejnew})-(\ref{eq:Ve3Uejnew}) may already be suggestive of the way to find out solutions of $\{z^{}_{+i}, z^{}_{-i}\}$ (for $i = 1, 2, 3$). On the other hand, we have seen that the polynomial invariants under the symmetry group $S^{}_3$ appear frequently in the differential equations, so the implementation of the invariant theory of finite groups might be helpful~\cite{Benson, Smith, Neusel}.

\section{Summary}
\label{sec:sum}

In this paper, we examine the continuous symmetries of the RGEs of effective neutrino mixing parameters in both cases of two- and three-flavor neutrino oscillations in matter. We have shown that two one-parameter Lie groups associated with the translational and scaling invariance play an important role and lead to explicit solutions by quadratures alone. In practice, the discrete symmetries, $Z^{}_2$ in the two-flavor case and $S^{}_3$ in the three-flavor case, could guide us to define new variables that belong to the irreducible representations of the symmetry group and help reduce the number of ordinary differential equations.

In the two-flavor case, the RGEs of two independent variables $\{|V^{}_{e1}|^2, \widetilde{\Delta}^{}_{21}\}$ are reformulated as those of $\{x^{}_+, x^{}_-\}$, which transform as the trivial and nontrivial one-dimensional representations of the $Z^{}_2$ group. As a result, the $Z^{}_2$ symmetry is maintained, whereas the symmetry is explicitly broken in the RGEs of $\{|V^{}_{e1}|^2, \widetilde{\Delta}^{}_{21}\}$. Furthermore, the translational symmetry generated by the vector field ${\bf v}^{}_1 \equiv \partial^{}_a$ and the scaling symmetry by ${\bf v}^{}_2 \equiv -a\partial^{}_a + x^{}_+\partial^{}_+ + x^{}_-\partial^{}_-$ guarantee that the RGEs of $\{x^{}_-, x^{}_+\}$ can be completely solved by quadratures. We present the explicit solutions in this case and prove that the differential invariant ${\cal I}(a, x^{}_+, x^{}_-) = (x^2_- - x^2_+)/x^4_- = 4 |V^{}_{e1}|^2 |V^{}_{e2}|^2 \widetilde{\Delta}^2_{21}$ is unique. Here the differential invariants refer to the invariant functions ${\cal I}(a, x^{}_+, x^{}_-)$ under the
group transformations generated by ${\bf v}^{}_1 \equiv \partial^{}_a$ and ${\bf w} \equiv ({\rm d}x^{}_+/{\rm d}a)\partial^{}_+ + ({\rm d}x^{}_-/{\rm d}a)\partial^{}_-$. Although the calculations in the two-flavor case are quite straightforward, we conclude with a negative answer to the question whether there are additional identities other than $|V^{}_{e1}||V^{}_{e2}| \widetilde{\Delta}^{}_{21} = |U^{}_{e1}||U^{}_{e2}| \Delta^{}_{21}$.

In the three-flavor case, instead of the four independent variables $\{|V^{}_{e1}|^2, |V^{}_{e2}|^2\}$ and $\{\widetilde{\Delta}^{}_{21}, \widetilde{\Delta}^{}_{31}\}$, which break the $S^{}_3$ symmetry, we introduce new variables $\{\widetilde{z}^{}_{+1}, \widetilde{z}^{}_{+2}, \widetilde{z}^{}_{+3}\}$ and $\{\widetilde{z}^{}_{-1}, \widetilde{z}^{}_{-2}, \widetilde{z}^{}_{-3}\}$ that transform as ${\bf 3} = {\bf 2} \oplus {\bf 1}$ and ${\bf 3}^\prime = {\bf 2} \oplus {\bf 1}^\prime$ under the $S^{}_3$ symmetry. By doing so, the $S^{}_3$ symmetry is preserved for the RGEs of $\{\widetilde{z}^{}_{\pm i}\}$ (for $i = 1, 2, 3$). With the help of two one-parameter Lie groups generated by ${\bf v}^{}_1 = \partial^{}_a$ and ${\bf v}^{}_2 \equiv -a\partial^{}_a + \widetilde{z}^{}_{+ i} \partial^{}_{+i} + \widetilde{z}^{}_{- i}\partial^{}_{-i}$, we have reduced the number of RGEs by two. However, the reduced set of four RGEs cannot be simply solved by quadratures. We argue that the number of differential invariants associated with two vector fields ${\bf v}^{}_1$ and ${\bf w} \equiv ({\rm d}\widetilde{z}^{}_{+i}/{\rm d}a)\partial^{}_{+i} + ({\rm d}\widetilde{z}^{}_{-i}/{\rm d}a)\partial^{}_{-i}$ should be one in the three-flavor case, however, an explicit construction is still lacking. In addition, the Naumov relation can be handled only when the RGEs are extended to include other elements of the effective mixing matrix, e.g., $|V^{}_{\mu i}|^2$ or $|V^{}_{\tau i}|^2$ for $i = 1, 2, 3$. Direct solutions to the extended set of differential equations will be more involved. However, the reformulation of the RGEs by following the symmetry principle may have already be useful in numerically solving the RGEs. A comparative study of numerical solutions to the RGEs in the conventional and symmetric formulation is interesting and will be carried out in a separate work.

The importance of both continuous and discrete symmetries can never be overemphasized. Further efforts in exploring the symmetries of the RGEs of effective neutrino mixing parameters and in solving directly the RGEs are desirable and could deepen greatly our understanding of the matter effects on neutrino oscillations. We believe that the results presented in the work are helpful for future works along this direction.

\section*{Acknowledgements}

The author would like to thank Di Zhang for a helpful discussion about the $S^{}_3$ symmetry. This work was supported in part by the National Natural Science Foundation of China under Grant No.~11775232 and No.~11835013, and by the CAS Center for Excellence in Particle Physics.

\appendix

\section{The $S^{}_3$ Symmetry Group}
\label{sec:app}

Discrete symmetry groups have been extensively studied in connection with fermion masses and flavor mixing~\cite{Ishimori:2010au, Ishimori:2012zz}. Among them, $S^{}_3$ is the simplest non-Abelian group and has attracted a lot of attention in the model building. In this section, we summarize the main features of the $S^{}_3$ group, and examine its reducible three-dimensional representation ${\bf 3}^\prime$, which has rarely been mentioned in the literature.

As is well known, the $S^{}_3$ group contains six elements $\{e, a, b, aba, ab, ba\}$, which can be divided into three conjugate classes ${\cal C}^{}_1 = \{e\}$, ${\cal C}^{}_2 = \{ab, ba\}$ and ${\cal C}^{}_3 = \{a, b, aba\}$, where two generators $a = S^{(12)}$ and $b = S^{(23)}$ have been identified. More explicitly, the reducible three-dimensional representation ${\bf 3}$ of the $S^{}_3$ group elements can be produced via
\begin{eqnarray}
\left(\begin{matrix} z^{}_{+1} \cr z^{}_{+2} \cr z^{}_{+3} \end{matrix}\right) & \xrightarrow[e]{I} & \left(\begin{matrix} 1 & 0 & 0 \cr 0 & 1 & 0 \cr 0 & 0 & 1\end{matrix}\right) \cdot \left(\begin{matrix} z^{}_{+1} \cr z^{}_{+2} \cr z^{}_{+3} \end{matrix}\right) = \left(\begin{matrix} z^{}_{+1} \cr z^{}_{+2} \cr z^{}_{+3} \end{matrix}\right) \; , \label{eq:S3e} \\
\left(\begin{matrix} z^{}_{+1} \cr z^{}_{+2} \cr z^{}_{+3} \end{matrix}\right) & \xrightarrow[a]{S^{(12)}} & \left(\begin{matrix} 0 & 1 & 0 \cr 1& 0 & 0 \cr 0 & 0 & 1\end{matrix}\right) \cdot \left(\begin{matrix} z^{}_{+1} \cr z^{}_{+2} \cr z^{}_{+3} \end{matrix}\right) = \left(\begin{matrix} z^{}_{+2} \cr z^{}_{+1} \cr z^{}_{+3} \end{matrix}\right) \; , \label{eq:S3a} \\
\left(\begin{matrix} z^{}_{+1} \cr z^{}_{+2} \cr z^{}_{+3} \end{matrix}\right) & \xrightarrow[b]{S^{(23)}} & \left(\begin{matrix} 1 & 0 & 0 \cr 0 & 0 & 1 \cr 0 & 1 & 0\end{matrix}\right) \cdot \left(\begin{matrix} z^{}_{+1} \cr z^{}_{+2} \cr z^{}_{+3} \end{matrix}\right) = \left(\begin{matrix} z^{}_{+1} \cr z^{}_{+3} \cr z^{}_{+2} \end{matrix}\right) \; , \label{eq:S3b} \\
\left(\begin{matrix} z^{}_{+1} \cr z^{}_{+2} \cr z^{}_{+3} \end{matrix}\right) & \xrightarrow[aba]{S^{(31)}} & \left(\begin{matrix} 0& 0 & 1 \cr 0 & 1 & 0 \cr 1 & 0 & 0\end{matrix}\right) \cdot \left(\begin{matrix} z^{}_{+1} \cr z^{}_{+2} \cr z^{}_{+3} \end{matrix}\right) = \left(\begin{matrix} z^{}_{+3} \cr z^{}_{+2} \cr z^{}_{+1} \end{matrix}\right) \; , \label{eq:S3aba} \\
\left(\begin{matrix} z^{}_{+1} \cr z^{}_{+2} \cr z^{}_{+3} \end{matrix}\right) & \xrightarrow[ab]{S^{(321)}} & \left(\begin{matrix} 0 & 0 & 1 \cr 1 & 0 & 0 \cr 0 & 1 & 0\end{matrix}\right) \cdot \left(\begin{matrix} z^{}_{+1} \cr z^{}_{+2} \cr z^{}_{+3} \end{matrix}\right) = \left(\begin{matrix} z^{}_{+3} \cr z^{}_{+1} \cr z^{}_{+2} \end{matrix}\right) \; , \label{eq:S3ab} \\
\left(\begin{matrix} z^{}_{+1} \cr z^{}_{+2} \cr z^{}_{+3} \end{matrix}\right) & \xrightarrow[ba]{S^{(123)}} & \left(\begin{matrix} 0& 1 & 0 \cr 0 & 0 & 1 \cr 1 & 0 & 0\end{matrix}\right) \cdot \left(\begin{matrix} z^{}_{+1} \cr z^{}_{+2} \cr z^{}_{+3} \end{matrix}\right) = \left(\begin{matrix} z^{}_{+2} \cr z^{}_{+3} \cr z^{}_{+1} \end{matrix}\right) \; , \label{eq:S3ba}
\end{eqnarray}
where the identification of the group elements with their representation matrices is evident. There are totally three irreducible representations for $S^{}_3$, namely, two one-dimensional representations ${\bf 1}$ and ${\bf 1}^\prime$, and one two-dimensional representation ${\bf 2}$. In the chosen basis $(z^{}_{+1}, z^{}_{+2}, z^{}_{+3})^{\rm T}$, we have constructed the reducible three-dimensional representation ${\bf 3}$, which can be decomposed into two irreducible representations as ${\bf 3} = {\bf 2} \oplus {\bf 1}$. This can be achieved via the basis transformation
\begin{eqnarray}
\left(\begin{matrix} \tilde{z}^{}_{+1} \cr \tilde{z}^{}_{+2} \cr \tilde{z}^{}_{+3}\end{matrix}\right) = \left(\begin{matrix} \displaystyle \frac{1}{\sqrt{2}} & \displaystyle -\frac{1}{\sqrt{2}} & 0 \cr \displaystyle \frac{1}{\sqrt{6}} & \displaystyle \frac{1}{\sqrt{6}} & \displaystyle -\frac{2}{\sqrt{6}} \cr \displaystyle \frac{1}{\sqrt{3}} & \displaystyle \frac{1}{\sqrt{3}} & \displaystyle \frac{1}{\sqrt{3}}\end{matrix}\right) \cdot \left(\begin{matrix} z^{}_{+1} \cr z^{}_{+2} \cr z^{}_{+3} \end{matrix}\right) \equiv R^{}_+ \left(\begin{matrix} z^{}_{+1} \cr z^{}_{+2} \cr z^{}_{+3} \end{matrix}\right) \; ,
\label{eq:S3T}
\end{eqnarray}
where the orthogonal matrix $R^{}_+$ can be used to derive the representation matrices of the $S^{}_3$ group elements in the new basis. It is straightforward to verify that the new representation matrix $S$ can be obtained via $\tilde{S} = R^{}_+ S R^{-1}_+$, where $S$ denotes any one of the representation matrices in Eqs.~(\ref{eq:S3e})-(\ref{eq:S3ba}). In the new basis $(\tilde{z}^{}_{+1}, \tilde{z}^{}_{+2}, \tilde{z}^{}_{+3})^{\rm T}$, we have
\begin{eqnarray}
\tilde{I} = \left(\begin{matrix} 1 & 0 & 0 \cr 0 & 1 & 0 \cr 0 & 0 & 1\end{matrix}\right) \; , \quad &~& \tilde{S}^{(12)} = \left(\begin{matrix} -1 & 0 & 0 \cr 0 & 1 & 0 \cr 0 & 0 & 1\end{matrix}\right) \; , \label{eq:S3r1} \\
\tilde{S}^{(23)} = \left(\begin{matrix} \displaystyle \frac{1}{2} & \displaystyle \frac{\sqrt{3}}{2} & 0 \cr \displaystyle \frac{\sqrt{3}}{2} & -\displaystyle \frac{1}{2} & 0 \cr 0 & 0 & 1\end{matrix}\right) \; , \quad &~& \tilde{S}^{(31)} = \left(\begin{matrix} \displaystyle \frac{1}{2} & \displaystyle -\frac{\sqrt{3}}{2} & 0 \cr \displaystyle -\frac{\sqrt{3}}{2} & -\displaystyle \frac{1}{2} & 0 \cr 0 & 0 & 1\end{matrix}\right) \; , \label{eq:S3r2} \\
\tilde{S}^{(321)} = \left(\begin{matrix} \displaystyle -\frac{1}{2} & \displaystyle -\frac{\sqrt{3}}{2} & 0 \cr \displaystyle \frac{\sqrt{3}}{2} & -\displaystyle \frac{1}{2} & 0 \cr 0 & 0 & 1\end{matrix}\right) \; , \quad &~& \tilde{S}^{(123)} = \left(\begin{matrix} \displaystyle -\frac{1}{2} & \displaystyle \frac{\sqrt{3}}{2} & 0 \cr \displaystyle -\frac{\sqrt{3}}{2} & -\displaystyle \frac{1}{2} & 0 \cr 0 & 0 & 1\end{matrix}\right) \; , \label{eq:S3r3}
\end{eqnarray}
which are block-diagonal. Therefore, starting with the triplet $(z^{}_{+1}, z^{}_{+2}, z^{}_{+3})^{\rm T}$, we have decomposed it into one doublet and one singlet, namely,
\begin{eqnarray}
D^{}_+ = \left(\begin{matrix} \tilde{z}^{}_{+1} \cr \tilde{z}^{}_{+2} \end{matrix}\right) = \left(\begin{matrix} \displaystyle \frac{1}{\sqrt{2}} \left(z^{}_{+1} - z^{}_{+2}\right) \cr \displaystyle \frac{1}{\sqrt{6}} \left(z^{}_{+1} + z^{}_{+2} - 2 z^{}_{+3}\right) \end{matrix}\right) \; , \quad S^{}_+ = \tilde{z}^{}_{+3} = \frac{1}{\sqrt{3}} \left(z^{}_{+1} + z^{}_{+2} + z^{}_{+3}\right) \; , \quad \label{eq:S3DS}
\end{eqnarray}
for which the transformation rules under the $S^{}_3$ group can be read off from Eqs.~(\ref{eq:S3r1})-(\ref{eq:S3r3}). In addition, the nontrivial reduction of the tensor products of any two irreducible representations can be summarized as below~\cite{Ishimori:2012zz}
\begin{eqnarray}
\left(\begin{matrix} w^{}_1 \cr w^{}_2 \end{matrix}\right)^{}_{\bf 2} \otimes \left(\begin{matrix} v^{}_1 \cr v^{}_2\end{matrix}\right)^{}_{\bf 2} = \left(\begin{matrix} w^{}_1 v^{}_2 + w^{}_2 v^{}_1 \cr w^{}_1 v^{}_1 - w^{}_2 v^{}_2\end{matrix}\right)^{}_{\bf 2} \oplus \left(w^{}_1 v^{}_2 - w^{}_2 v^{}_1\right)^{}_{{\bf 1}^\prime} \oplus \left(w^{}_1 v^{}_1 + w^{}_2 v^{}_2\right)^{}_{\bf 1} \; , \label{eq:S3wv}
\end{eqnarray}
and
\begin{eqnarray}
\left(\begin{matrix} w^{}_1 \cr w^{}_2\end{matrix}\right)^{}_{\bf 2} \otimes \left(v\right)^{}_{\bf 1} = \left(\begin{matrix} w^{}_1 v \cr w^{}_2 v\end{matrix}\right)^{}_{\bf 2} \; , \quad \left(\begin{matrix} w^{}_1 \cr w^{}_2\end{matrix}\right)^{}_{\bf 2} \otimes \left(v^\prime\right)^{}_{{\bf 1}^\prime} = \left(\begin{matrix} -w^{}_2 v^\prime \cr +w^{}_1 v^\prime\end{matrix}\right)^{}_{\bf 2} \; , \label{eq:S3wv2}
\end{eqnarray}
where the doublet is involved, while $\left(w\right)^{}_{\bf 1} \otimes \left(v^\prime\right)^{}_{{\bf 1}^\prime} = \left(wv^\prime\right)^{}_{{\bf 1}^\prime}$ and $\left(w^\prime\right)^{}_{{\bf 1}^\prime} \otimes \left(v^\prime\right)^{}_{{\bf 1}^\prime} = \left(w^\prime v^\prime\right)^{}_{\bf 1}$ for singlets.

Then we turn to another reducible three-dimensional representation ${\bf 3}^\prime$, which will be represented by $(z^{}_{-1}, z^{}_{-2}, z^{}_{-3})^{\rm T}$, and the corresponding transformations under the $S^{}_3$ group read
\begin{eqnarray}
\left(\begin{matrix} z^{}_{-1} \cr z^{}_{-2} \cr z^{}_{-3} \end{matrix}\right) & \xrightarrow[e]{I} & \left(\begin{matrix} 1 & 0 & 0 \cr 0 & 1 & 0 \cr 0 & 0 & 1\end{matrix}\right) \cdot \left(\begin{matrix} z^{}_{-1} \cr z^{}_{-2} \cr z^{}_{-3} \end{matrix}\right) = \left(\begin{matrix} z^{}_{-1} \cr z^{}_{-2} \cr z^{}_{-3} \end{matrix}\right) \; , \label{eq:S3pe} \\
\left(\begin{matrix} z^{}_{-1} \cr z^{}_{-2} \cr z^{}_{-3} \end{matrix}\right) & \xrightarrow[a]{S^{(12)}} & \left(\begin{matrix} 0 & -1 & 0 \cr -1 & 0 & 0 \cr 0 & 0 & -1\end{matrix}\right) \cdot \left(\begin{matrix} z^{}_{-1} \cr z^{}_{-2} \cr z^{}_{-3} \end{matrix}\right) = \left(\begin{matrix} -z^{}_{-2} \cr -z^{}_{-1} \cr -z^{}_{-3} \end{matrix}\right) \; , \label{eq:S3pa} \\
\left(\begin{matrix} z^{}_{-1} \cr z^{}_{-2} \cr z^{}_{-3} \end{matrix}\right) & \xrightarrow[b]{S^{(23)}} & \left(\begin{matrix} -1 & 0 & 0 \cr 0 & 0 & -1 \cr 0 & -1 & 0\end{matrix}\right) \cdot \left(\begin{matrix} z^{}_{-1} \cr z^{}_{-2} \cr z^{}_{-3} \end{matrix}\right) = \left(\begin{matrix} -z^{}_{-1} \cr -z^{}_{-3} \cr -z^{}_{-2} \end{matrix}\right) \; , \label{eq:S3pb} \\
\left(\begin{matrix} z^{}_{-1} \cr z^{}_{-2} \cr z^{}_{-3} \end{matrix}\right) & \xrightarrow[aba]{S^{(31)}} & \left(\begin{matrix} 0& 0 & -1 \cr 0 & -1 & 0 \cr -1 & 0 & 0\end{matrix}\right) \cdot \left(\begin{matrix} z^{}_{-1} \cr z^{}_{-2} \cr z^{}_{-3} \end{matrix}\right) = \left(\begin{matrix} -z^{}_{-3} \cr -z^{}_{-2} \cr -z^{}_{-1} \end{matrix}\right) \; , \label{eq:S3paba} \\
\left(\begin{matrix} z^{}_{-1} \cr z^{}_{-2} \cr z^{}_{-3} \end{matrix}\right) & \xrightarrow[ab]{S^{(321)}} & \left(\begin{matrix} 0 & 0 & 1 \cr 1 & 0 & 0 \cr 0 & 1 & 0\end{matrix}\right) \cdot \left(\begin{matrix} z^{}_{-1} \cr z^{}_{-2} \cr z^{}_{-3} \end{matrix}\right) = \left(\begin{matrix} z^{}_{-3} \cr z^{}_{-1} \cr z^{}_{-2} \end{matrix}\right) \; , \label{eq:S3pab} \\
\left(\begin{matrix} z^{}_{-1} \cr z^{}_{-2} \cr z^{}_{-3} \end{matrix}\right) & \xrightarrow[ba]{S^{(123)}} & \left(\begin{matrix} 0& 1 & 0 \cr 0 & 0 & 1 \cr 1 & 0 & 0\end{matrix}\right) \cdot \left(\begin{matrix} z^{}_{-1} \cr z^{}_{-2} \cr z^{}_{-3} \end{matrix}\right) = \left(\begin{matrix} z^{}_{-2} \cr z^{}_{-3} \cr z^{}_{-1} \end{matrix}\right) \; . \label{eq:S3pba}
\end{eqnarray}
In the differential equations considered in the main text, both $(z^{}_{+1}, z^{}_{+2}, z^{}_{+3})^{\rm T}$ and $(z^{}_{-1}, z^{}_{-2}, z^{}_{-3})^{\rm T}$ are present, which transform as ${\bf 3}$ and ${\bf 3}^\prime$ under the $S^{}_3$ group, respectively. Obviously, this three-dimensional representation can in principle be decomposed into two irreducible representations, i.e., ${\bf 3}^\prime = {\bf 2} \oplus {\bf 1}^\prime$. Such a decomposition can be carried out in two steps.

First, since the characters of the nontrivial one-dimensional representation ${\bf 1}^\prime$ for three conjugate classes are given by $\chi^{}_{{\bf 1}^\prime}({\cal C}^{}_1) = 1$, $\chi^{}_{{\bf 1}^\prime}({\cal C}^{}_2) = 1$ and $\chi^{}_{{\bf 1}^\prime}({\cal C}^{}_3) = -1$, they constitute the one-dimensional matrix representations of the group elements in each conjugate class. In the same way as for the triplet ${\bf 3}$, we can construct the singlet ${\bf 1}^\prime$ from the triplet ${\bf 3}^\prime$ through
\begin{eqnarray}
S^{}_- = \tilde{z}^{}_{-3} \equiv \frac{1}{\sqrt{3}} \left(z^{}_{-1} + z^{}_{-2} + z^{}_{-3}\right) \; . \label{eq:S3pS}
\end{eqnarray}
Under those transformations given in Eqs.~(\ref{eq:S3pe})-(\ref{eq:S3pba}), one can immediately verify that $S^{}_-$ belongs to the nontrivial one-dimensional representation ${\bf 1}^\prime$.

Then, we introduce a new triplet $(z^\prime_{-1}, z^\prime_{-2}, z^\prime_{-3})^{\rm T}$, which is actually built upon the original one $(z^{}_{-1}, z^{}_{-2}, z^{}_{-3})^{\rm T}$ via
\begin{eqnarray}
z^\prime_{-1} \equiv z^{}_{-2} - z^{}_{-3} \; , \quad z^\prime_{-2} \equiv z^{}_{-3} - z^{}_{-1} \; , \quad z^\prime_{-3} \equiv z^{}_{-1} - z^{}_{-2} \; .
\label{eq:zprime}
\end{eqnarray}
Notice that only two components of $(z^\prime_{-1}, z^\prime_{-2}, z^\prime_{-3})^{\rm T}$ can actually be independent due to the identity $z^\prime_{-1} + z^\prime_{-2} + z^\prime_{-3} = 0$. Furthermore, one can easily check that $(z^\prime_{-1}, z^\prime_{-2}, z^\prime_{-3})^{\rm T}$ transforms as ${\bf 3}^\prime$, from which the following doublet can be found
\begin{eqnarray}
D^\prime_- \equiv \left(\begin{matrix} \displaystyle \frac{1}{\sqrt{2}} \left(z^\prime_{-1} - z^\prime_{-2}\right) \cr \displaystyle \frac{1}{\sqrt{6}} \left(z^\prime_{-1} + z^\prime_{-2} - 2z^\prime_{-3}\right) \end{matrix}\right) = \sqrt{3} \left( \begin{matrix} \displaystyle \frac{1}{\sqrt{6}} \left(z^{}_{-1} + z^{}_{-2} - 2 z^{}_{-3}\right) \cr \displaystyle -\frac{1}{\sqrt{2}} \left(z^{}_{-1} - z^{}_{-2}\right) \end{matrix} \right) \; ,
\label{eq:S3pDp}
\end{eqnarray}
where the definitions of $z^\prime_{-i}$ for $i = 1, 2, 3$ in Eq.~(\ref{eq:zprime}) have been implemented. Consequently, we can choose the basis
\begin{eqnarray}
\left(\begin{matrix} \tilde{z}^{}_{-1} \cr \tilde{z}^{}_{-2} \cr \tilde{z}^{}_{-3}\end{matrix}\right) = \left(\begin{matrix} \displaystyle \frac{1}{\sqrt{6}} & \displaystyle \frac{1}{\sqrt{6}} & \displaystyle -\frac{2}{\sqrt{6}} \cr \displaystyle -\frac{1}{\sqrt{2}} & \displaystyle \frac{1}{\sqrt{2}} & 0 \cr \displaystyle \frac{1}{\sqrt{3}} & \displaystyle \frac{1}{\sqrt{3}} & \displaystyle \frac{1}{\sqrt{3}}\end{matrix}\right) \cdot \left(\begin{matrix} z^{}_{-1} \cr z^{}_{-2} \cr z^{}_{-3} \end{matrix}\right) \equiv R^{}_- \left(\begin{matrix} z^{}_{-1} \cr z^{}_{-2} \cr z^{}_{-3} \end{matrix}\right) \; ,
\label{eq:S3P}
\end{eqnarray}
leading to the reduction ${\bf 3}^\prime = {\bf 2} \oplus {\bf 1}^\prime$ with the doublet
\begin{eqnarray}
D^{}_- \equiv \left(\begin{matrix} \tilde{z}^{}_{-1} \cr \tilde{z}^{}_{-2}\end{matrix}\right) = \left( \begin{matrix} \displaystyle \frac{1}{\sqrt{6}} \left(z^{}_{-1} + z^{}_{-2} - 2 z^{}_{-3}\right) \cr \displaystyle -\frac{1}{\sqrt{2}} \left(z^{}_{-1} - z^{}_{-2}\right) \end{matrix} \right) \; ,
\label{eq:S3pD}
\end{eqnarray}
and the singlet $S^{}_-$ in Eq.~(\ref{eq:S3pS}). In a similar way, one can calculate the representation matrices in Eqs.~(\ref{eq:S3pe})-(\ref{eq:S3pba}) in the new basis via the similarity transformation with the orthogonal matrix $R^{}_-$ in Eq.~(\ref{eq:S3P}). The final results are
\begin{eqnarray}
\tilde{I} = \left(\begin{matrix} 1 & 0 & 0 \cr 0 & 1 & 0 \cr 0 & 0 & 1\end{matrix}\right) \; , \quad &~& \tilde{S}^{(12)\prime} = \left(\begin{matrix} -1 & 0 & 0 \cr 0 & 1 & 0 \cr 0 & 0 & -1\end{matrix}\right) \; , \label{eq:S3pr1} \\
\tilde{S}^{(23)\prime} = \left(\begin{matrix} \displaystyle \frac{1}{2} & \displaystyle \frac{\sqrt{3}}{2} & 0 \cr \displaystyle \frac{\sqrt{3}}{2} & -\displaystyle \frac{1}{2} & 0 \cr 0 & 0 & -1\end{matrix}\right) \; , \quad &~& \tilde{S}^{(31)\prime} = \left(\begin{matrix} \displaystyle \frac{1}{2} & \displaystyle -\frac{\sqrt{3}}{2} & 0 \cr \displaystyle -\frac{\sqrt{3}}{2} & -\displaystyle \frac{1}{2} & 0 \cr 0 & 0 & -1\end{matrix}\right) \; , \label{eq:S3pr2} \\
\tilde{S}^{(321)\prime} = \left(\begin{matrix} \displaystyle -\frac{1}{2} & \displaystyle -\frac{\sqrt{3}}{2} & 0 \cr \displaystyle \frac{\sqrt{3}}{2} & -\displaystyle \frac{1}{2} & 0 \cr 0 & 0 & 1\end{matrix}\right) \; , \quad &~& \tilde{S}^{(123)\prime} = \left(\begin{matrix} \displaystyle -\frac{1}{2} & \displaystyle \frac{\sqrt{3}}{2} & 0 \cr \displaystyle -\frac{\sqrt{3}}{2} & -\displaystyle \frac{1}{2} & 0 \cr 0 & 0 & 1\end{matrix}\right) \; , \label{eq:S3pr3}
\end{eqnarray}
from which one can see that the upper-left $2\times 2$ blocks are exactly the same as their counterparts in Eqs.~(\ref{eq:S3r1})-(\ref{eq:S3r3}). This demonstrates that $D^{}_-$ in Eq.~(\ref{eq:S3pD}) and $S^{}_-$ in Eq.~(\ref{eq:S3pS}) transform as ${\bf 2}$ and ${\bf 1}^\prime$, respectively.

As we have seen from the previous discussions, both $D^{}_+$ and $D^{}_-$ are doublets, while $S^{}_+$ and $S^{}_-$ are respectively the trivial and nontrivial singlets, under the $S^{}_3$ symmetry. By the number of powers of $\widetilde{z}^{}_{+i}$ or $\widetilde{z}^{}_{-i}$ (for $i = 1, 2, 3$), the $S^{}_3$ invariants can be classified as follows.
\begin{itemize}
\item As we have demonstrated, $S^{}_+$ transforms as a singlet, while $S^{}_-$ as a nontrivial singlet. Therefore, the invariant linear in $\widetilde{z}^{}_{+i}$ or $\widetilde{z}^{}_{-i}$ is $S^{}_+ = \widetilde{z}^{}_{+3}$, which is unique at this level.

\item Then we consider the invariants quadratic in $z^{}_{+i}$ or $z^{}_{-i}$, which can be constructed from $D^{}_\pm$ and $S^{}_\pm$, namely, $S^2_+ = \widetilde{z}^2_{+3}$, $S^2_- = \widetilde{z}^2_{-3}$ and
\begin{eqnarray}
\left(D^{}_+ \otimes D^{}_+\right)^{}_{\bf 1} &=& \widetilde{z}^2_{+1} + \widetilde{z}^2_{+2} \; , \nonumber \\
\left(D^{}_- \otimes D^{}_-\right)^{}_{\bf 1} &=& \widetilde{z}^2_{-1} + \widetilde{z}^2_{-2} \; , \\
\left(D^{}_+ \otimes D^{}_-\right)^{}_{\bf 1} &=& \widetilde{z}^{}_{+1} \widetilde{z}^{}_{-1} + \widetilde{z}^{}_{+2} \widetilde{z}^{}_{-2} \; . \nonumber
\end{eqnarray}
In addition, $S^{}_+ S^{}_- = \widetilde{z}^{}_{+3} \widetilde{z}^{}_{-3}$ and $\left(D^{}_+ \otimes D^{}_-\right)^{}_{{\bf 1}^\prime} = \widetilde{z}^{}_{+1} \widetilde{z}^{}_{-2} - \widetilde{z}^{}_{+2} \widetilde{z}^{}_{-1}$ will receive an extra minus sign under the transformations in the conjugate class ${\cal C}^{}_3 =\{a, b, aba\}$.

\item For the invariants of the third power of $\widetilde{z}^{}_{+i}$ or $\widetilde{z}^{}_{-i}$, we have $S^3_+ = \widetilde{z}^3_{+3}$, $S^{}_+ S^2_- = \widetilde{z}^{}_{+3} \widetilde{z}^2_{-3}$, and
\begin{eqnarray}
\left(D^{}_+ \otimes D^{}_+\right)^{}_{\bf 1} S^{}_+ &=& (\widetilde{z}^2_{+1} + \widetilde{z}^2_{+2}) \widetilde{z}^{}_{+3} \; , \nonumber \\
\left(D^{}_- \otimes D^{}_-\right)^{}_{\bf 1} S^{}_+ &=& (\widetilde{z}^2_{-1} + \widetilde{z}^2_{-2}) \widetilde{z}^{}_{+3} \; , \nonumber \\
\left(D^{}_+ \otimes D^{}_-\right)^{}_{\bf 1} S^{}_+ &=& (\widetilde{z}^{}_{+1} \widetilde{z}^{}_{-1} + \widetilde{z}^{}_{+2} \widetilde{z}^{}_{-2})\widetilde{z}^{}_{-3} \; , \nonumber \\
\left(D^{}_+ \otimes D^{}_-\right)^{}_{{\bf 1}^\prime} S^{}_- &=& (\widetilde{z}^{}_{+1} \widetilde{z}^{}_{-2} - \widetilde{z}^{}_{+2} \widetilde{z}^{}_{-1})\widetilde{z}^{}_{-3} \; , \nonumber \\
\left(D^{}_+ \otimes D^{}_+ \otimes D^{}_-\right)^{}_{\bf 1} &=& 2 \widetilde{z}^{}_{+ 1} \widetilde{z}^{}_{+ 2} \widetilde{z}^{}_{- 1} + (\widetilde{z}^2_{+ 1} - \widetilde{z}^2_{+ 2}) \widetilde{z}^{}_{- 2} \; , \\
\left(D^{}_- \otimes D^{}_- \otimes D^{}_+\right)^{}_{\bf 1} &=& 2 \widetilde{z}^{}_{- 1} \widetilde{z}^{}_{- 2} \widetilde{z}^{}_{+ 1} + (\widetilde{z}^2_{- 1} - \widetilde{z}^2_{- 2}) \widetilde{z}^{}_{+ 2} \; , \nonumber \\
\left(D^{}_+ \otimes D^{}_+ \otimes D^{}_+\right)^{}_{\bf 1} &=& 3\widetilde{z}^2_{+1} \widetilde{z}^{}_{+2} - \widetilde{z}^3_{+ 2} \;, \nonumber \\
\left(D^{}_- \otimes D^{}_- \otimes D^{}_- \right)^{}_{\bf 1} &=& 3\widetilde{z}^2_{- 1} \widetilde{z}^{}_{-2} - \widetilde{z}^3_{- 2} \; . \nonumber
\end{eqnarray}
\end{itemize}
We can figure out all the invariants of higher orders in a similar way, but they are not useful for our discussions and will be omitted.

\end{document}